\documentclass[oldversion]{aa}  

\usepackage{graphicx}
\usepackage{txfonts}
\usepackage{float}
\usepackage[latin1]{inputenc}
\usepackage{supertabular}
\usepackage[normalem]{ulem}

\usepackage{natbib}
\bibpunct{(}{)}{;}{a}{}{,}

\newcommand{\FAC}{HE~1327$-$2326} 
\newcommand{\CBB}{HE~0107$-$5240} 
\newcommand{\cd}{CD~$-38^{\circ}\,245$}            
\newcommand{\teffm}{T_{\mbox{\scriptsize eff}}}    
\newcommand{\rb}[1]{\raisebox{1.5ex}[-1.5ex]{#1}}

\begin{document}

\title{The stellar content of the Hamburg/ESO survey\thanks{Based on
    observations collected at the European Southern Observatory, Chile
    (Proposal ID 145.B-0009). Tables A.1 and A.2 are only available in
    electronic form at the CDS via anonymous ftp to cdsarc.u-strasbg.fr
    (130.79.125.5) or via http://cdsweg.u-strasbg.fr/Abstract.html.}
      } 

\subtitle{IV. Selection of candidate metal-poor stars}

\author{
  N. Christlieb\inst{1,2} \and
  T. Sch\"orck\inst{2}\and
  A. Frebel\inst{3}\and
  T.C. Beers\inst{4} \and
  L. Wisotzki\inst{5} \and
  D. Reimers\inst{2}
}

\institute{
     Department of Astronomy and Space Physics, Uppsala University, Box 515, 
     75120 Uppsala, Sweden\\
     \email{norbert@astro.uu.se}
\and Hamburger Sternwarte, Universit\"at Hamburg, Gojenbergsweg 112,
     D-21029 Hamburg, Germany\\
     \email{nchristlieb/dreimers@hs.uni-hamburg.de}
\and McDonald Observatory, The University of Texas at Austin, 1 University
     Station, C1400, Austin, TX~78712-0259\\
     \email{anna@astro.as.utexas.edu}
\and Department of Physics and Astronomy, CSCE: Center for the Study of
     Cosmic Evolution, and JINA: Joint Institute for Nuclear Astrophysics,
     Michigan State University, E. Lansing, MI 48824, USA\\
     \email{beers@pa.msu.edu}
\and Astrophysical Institute Postsdam, An der Sternwarte 16,
     D-14482 Potsdam, Germany\\
     \email{lutz@aip.de}
}

\offprints{N. Christlieb,\\ \email{norbert@astro.uu.se}}
\date{Received 28 September 2007 / Accepted 4 March 2008}
\titlerunning{Metal-poor stars from the HES}
\authorrunning{Christlieb et al.}

\abstract{  
  
  We present the quantitative methods used for selecting candidate metal-poor
  stars in the Hamburg/ESO objective-prism survey (HES). The selection is
  based on the strength of the \ion{Ca}{ii}~K line, $B-V$ colors (both
  measured directly from the digital HES spectra), as well as $J-K$ colors
  from the 2 Micron All Sky Survey. The KP index for \ion{Ca}{ii}~K can be
  measured from the HES spectra with an accuracy of 1.0\,{\AA}, and a
  calibration of the HES $B-V$ colors, using CCD photometry, yields a
  1-$\sigma$ uncertainty of $0.07$\,mag for stars in the color range $0.3 <
  B-V < 1.4$. These accuracies make it possible to reliably reject stars with
  $\mbox{[Fe/H]} > -2.0$ without sacrificing completeness at the lowest
  metallicities. A test of the selection using 1121 stars of the HK survey of
  Beers, Preston, and Shectman present on HES plates suggests that the
  completeness at $\mbox{[Fe/H]} < -3.5$ is close to 100\,\% and that, at the
  same time, the contamination of the candidate sample with false positives is
  low: 50\,\% of all stars with $\mbox{[Fe/H]} > -2.5$ and 97\,\% of all stars
  with $\mbox{[Fe/H]} > -2.0$ are rejected. The selection was applied to 379
  HES fields, covering a nominal area of $8853$\,deg$^2$ of the southern high
  Galactic latitude sky. The candidate sample consists of 20,271 stars in the
  magnitude range $10 \lesssim B \lesssim 18$. A comparison of the magnitude
  distribution with that of the HK survey shows that the magnitude limit of
  the HES sample is about 2\,mag fainter. Taking the overlap of the sky areas
  covered by both surveys into account, it follows that the survey volume for
  metal-poor stars has been increased by the HES by about a factor of 10 with
  respect to the HK survey. We have already identified several very rare
  objects with the HES, including, e.g., the three most heavy-element
  deficient stars currently known.  }


\maketitle

\section{Introduction}\label{Sect:intro}

The chemical abundances of the atmospheres of metal-poor stars preserve, to a
large extent, the chemical composition of the gas clouds from which they
formed.  Therefore, metal-poor (and thus old) stars provide an observational
channel through which we can study the early history of the Galaxy and the
Universe.  These stars can be used, e.g., for determining a lower limit for
the age of the Universe, estimating the amount of \element[][7]{Li} produced
in Big Bang Nucleosynthesis, or constraining the elemental yields of the first
generations of supernovae (SNe). Details on these and other topics relevant to
the uses of metal-poor stars can be found in the review by
\citet{Beers/Christlieb:2005}.

The Hamburg/ESO objective-prism Survey (HES) was originally conceived for
finding bright quasars \citep{Reimers:1990,hespaperI,hespaperIII}. It was
carried out as an ESO Key Program (proposal 145.B-0009; P.I.: D. Reimers). For
details of the concept behind the survey and the data processing, we refer the
interested reader to \citet{hespaperIII}. The data quality of the HES offered
an efficient means of exploiting the stellar content of the survey. In
previous papers of this series, we reported on searches for DA white dwarfs
(\citealt{HESStarsI}; Paper~I), high-latitude carbon stars
(\citealt{HESStarsII}; Paper~II), and field horizontal-branch stars
(\citealt{HESFHBA}; Paper~III). In this paper we describe quantitative methods
for selecting candidates for very metal-poor stars (i.e., stars at
$\mbox{[Fe/H]} < -2.5$) in the HES database of digital objective-prism
spectra.

After a brief description of the HES (Sect. \ref{Sect:HESdescription}) we
present calibrations of line indices, an improved calibration of $B-V$ colors
estimated directly from the HES spectra (Sect.  \ref{Sect:Calibrations}). The
quantitative criteria that were applied to the digital HES spectra for
selecting metal-poor candidates, as well as the visual inspection procedure
for the selected candidates, are described in Sect. \ref{Sect:Selection}. An
evaluation of the selection is presented in Sect. \ref{Sect:Evaluation}. In
Sect. \ref{Sect:Results} we report on some of the basic properties of the
metal-poor candidate sample, including the magnitude distribution, before we
present our conclusions in Sect. \ref{Sect:Conclusions}.

Results of spectroscopic follow-up observations of the candidates will be
presented in forthcoming papers. However, the results for bright candidates
from 329 HES plates have already been reported in \citet{Frebeletal:2006b},
and the results for fainter stars have occasionally been published in papers
focusing on abundance analyses of HES metal-poor stars based on
high-resolution spectroscopy (see Sect.~\ref{Sect:Highres} for references).
We note that CCD photometry for many of the most interesting metal-poor
candidates has already been published in \citet{Beersetal:2007}.

\section{The Hamburg/ESO Survey}\label{Sect:HESdescription}

\subsection{Basic survey properties}\label{Sect:HESproperties}

The HES is based on photographic plates taken with the 1\,m ESO Schmidt
telescope, using its 4$^{\circ}$ prism. The 379 HES fields cover a total
nominal area of $8853\,\mathrm{deg}^2$ of the southern high Galactic latitude
sky. Taking overlapping plates and plate quarters as well as losses in area
due to overlapping spectra into account, the effective survey area is
$6726\,\mathrm{deg}^2$. The plates have been scanned at Hamburger Sternwarte
with a PDS microdensitometer. The large physical size of the ESO Schmidt
plates required that each plate be scanned in four separate sections, so that
the full matrix scans of each HES plate consists of four parts of $7500\times
7500$ pixels each.

The input catalog for extracting the HES spectra was generated from direct
images of the Digitized Sky Survey-I (DSS-I).\footnote{The DSS-I is based on
  photographic data obtained using The UK Schmidt Telescope. The UK Schmidt
  Telescope was operated by the Royal Observatory Edinburgh, with funding from
  the UK Science and Engineering Research Council, until 1988 June, and
  thereafter by the Anglo-Australian Observatory. Original plate material is
  copyright (c) the Royal Observatory Edinburgh and the Anglo-Australian
  Observatory. The plates were processed into the present compressed digital
  form with their permission. The Digitized Sky Survey was produced at the
  Space Telescope Science Institute under US Government grant NAG W-2166.}
Each of the objects in the input catalog was assigned to one of the following
three object classes: unsaturated point sources (source type
``\texttt{stars}''), bright sources above a saturation threshold determined by
examining the characteristic curve of each individual HES plate
(``\texttt{bright}''), and objects that have been classified as extended
sources by an automated morphological classification based on the DSS-I images
(``\texttt{ext}''). Note that the last class of sources does not exclusively
consist of galaxies and other objects that are indeed extended, but also
point-like sources located in the diffraction spikes of very bright stars, or
pairs of objects located very close to one another that are not separated on
the DSS-I direct image.

A photometric calibration of the HES was established through photometric
sequences obtained for all fields, mainly in the course of the Reimers et al.
ESO Key Program, and augmented by sequences taken from the latest version of
the \emph{Guide Star Photometric Catalog 2} \citep{Bucciarellietal:2001}. The
overall 1-$\sigma$ accuracy, including zero point errors, is better than
$0.15$\,mag in $B_J$ in the majority of the fields, degrading to $0.20$\,mag
in a few fields in which only sequences of lower quality are available.

An astrometric transformation between the DSS-I and HES plates was determined,
yielding for each object in the input catalog the $x,y$ position on the HES
plate used for extracting its spectrum in the sky-background reduced HES plate
scan. The HES spectra are extracted by algorithms optimized for each of the
above mentioned source types (see \citealt{hespaperIII} for details). The HES
data base consists of 12,357,153 digital spectra extracted on 379 plates. This
data base, as well as the full-matrix HES plate scans, will be made available
online in the near future.

\begin{figure}[htbp]
  \centering
  \includegraphics[clip=true,bb=54 597 442 778,width=9cm]{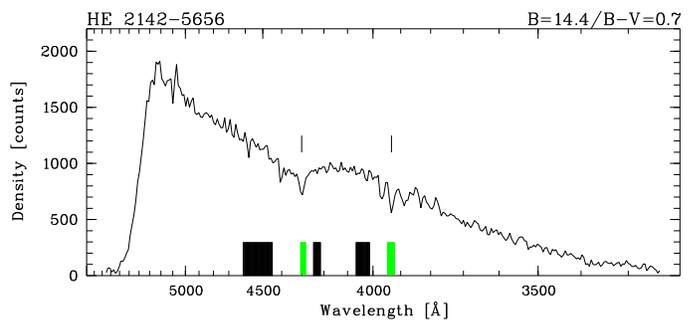}
  \caption{\label{Fig:HESspectrum_indices} HES example spectrum. The positions
    of the G~band of CH and the \ion{Ca}{ii}~K lines are marked, and the
    wavelength regions of the continuum (black) and line (grey) passbands for
    the measurement of the GP and KP indices are indicated.}
\end{figure}

The wavelength coverage of the HES spectra is limited by the atmospheric
cutoff at the blue end, and the sharp sensitivity cutoff of the IIIa-J
emulsion (``red edge''), resulting in a wavelength range of $3200\,\mbox{\AA}
< \lambda < 5300\,\mbox{\AA}$ (see Fig. \ref{Fig:HESspectrum_indices}). The
spectral resolution of the HES is mainly seeing-limited; it is typically
$\Delta\lambda = 10$\,{\AA} at the location of the \ion{Ca}{ii}~K line, at
$\lambda = 3934$\,{\AA}.

The quality of the HES spectra make this survey well-suited for executing a
search for metal-poor stars. In particular, as can be seen in
Fig.~\ref{Fig:HESmphs_examples}, the spectral resolution is sufficient to
detect the \ion{Ca}{ii}~K line even in quite metal-poor stars. This line can
be used as an indicator for [Fe/H], since for the overwhelming majority of the
stars in the Galaxy, the abundance ratio [Ca/Fe] follows a well-defined trend.
The spectral coverage of the HES spectra is broad enough to provide some color
information directly from the objective-prism spectra. Even though the $V$
band is not fully covered by the HES spectra, there is a strong correlation
between a so-called ``half power point'' -- a bisecting point of the
photographic density distribution in an appropriate wavelength range (see
Fig.~7 of \citealt{HESStarsI} for an illustration) -- and the $B-V$ color.
\citet{HESStarsI} showed that this correlation can be used to estimate $B-V$
colors directly from the HES spectra with an accuracy of $\sim 0.1$\,mag.
Color information is important for avoiding a bias against cool metal-poor
giants, which would occur if the stars were selected based only on their
\ion{Ca}{ii}~K line strength, because in cool giants this line has a
considerable strength even if the stars are metal poor.

With a limiting magnitude for stellar work of $B\sim 17.5$, another advantage
of the HES is that it is $\sim 2$\,mag deeper than the previously largest
survey for metal-poor stars, the so-called HK survey of Beers, Preston, and
Shectman \citep{BPSI,BPSII}. Taking the overlap in area of the HK and HE
surveys into account, it follows that with the HES, the total survey volume
for metal-poor stars is increased by about a factor of $\sim 10$ with respect
to the HK survey alone.

\subsection{Follow-up spectroscopy of metal-poor candidates}

As a result of the increased survey volume, the ongoing spectroscopic
follow-up observations of candidate metal-poor stars identified in the HES
have led to the discovery of stars that could not have been found in previous
surveys, due to their extreme rarity. For example, the three most
heavy-element deficient stars currently known were found in the HES: {\FAC}
($\mathrm{[Fe/H]} = -5.4$;
\citealt{Frebeletal:2005,Aokietal:2006,Frebeletal:2006a})\footnote{$\mbox{[X/H]}
  = \log_{10}\mbox{[$N$(X)/$N$(H)]}_{\ast} -
  \log_{10}\mbox{[$N$(X)/$N$(H)]}_{\sun}$, and analogously for [X/Fe]. $N$(X)
  is the number density of atoms of the element X.}; {\CBB} ($\mathrm{[Fe/H]}
= -5.7$;
\citealt{HE0107_Nature,HE0107_ApJ,Besselletal:2004,Christliebetal:2008b}); and
HE~0557$-$4840 ($\mathrm{[Fe/H]} = -4.7$; \citealt{Norrisetal:2007}). That is,
some 20 years after \citet{Bessell/Norris:1984} discovered the extremely
metal-poor giant {\cd}, with ${\rm [Fe/H]} \sim -4.0$
\citep{Norrisetal:2000,Francoisetal:2003}, it is now possible to study, by
means of long-lived stars, the history of our Galaxy when it was enriched by
heavy-elements to a level of only $\mathrm{[Fe/H]} \sim -5.0$, corresponding
to redshifts of $z > 5$ \citep{Clarke/Bromm:2003}.  It can be expected that
new, even deeper surveys for metal-poor stars, such as SEGUE: The Sloan
Extension for Galactic Understanding and
Exploration\footnote{\texttt{www.sdss.org/segue}}, will yield significant
numbers of additional stars in this [Fe/H] range. A review of past, present,
and future metal-poor star surveys can be found in
\citet{Beers/Christlieb:2005}.

\subsection{High-resolution spectroscopy of confirmed metal-poor stars}\label{Sect:Highres}

In addition to the objects mentioned in the previous section, many confirmed
metal-poor stars from the HES have already been observed at high spectral
resolution (i.e., $R = \lambda/\Delta\lambda \ge 40,000$). The spectra were
obtained at 8\,m-class telescopes, including Keck-I
\citep{KeckpaperI,KeckpaperII,KeckpaperIII,Cohenetal:2003,KeckpaperIV,Cohenetal:2005,Cohenetal:2006},
VLT-UT2 \citep{Depagneetal:2000,Frebeletal:2007b}, Subaru
\citep{Goswamietal:2006,Frebeletal:2007a,Aokietal:2007}, and Magellan-I.

A dedicated observational program for identifying metal-poor stars that are
strongly enhanced in r-process elements, the Hamburg/ESO R-process Enhanced
star Survey (HERES; \citealt{HERESpaperI}), has also been carried out with the
VLT. So-called ``snapshot spectra'' (i.e., spectra with $R\sim 20,000$ and
$S/N\sim 50$ per pixel) of 373 metal-poor stars, of which 346 stem from the
HES, were obtained with VLT/UVES. \citet{HERESpaperII} report on the results
of an automated and homogeneous abundance analysis of 253 stars whose spectra
are not heavily contaminated with CH lines. The carbon and nitrogen abundances
of 94 HERES stars, including 72 stars not analyzed by \citet{HERESpaperII}
because of their strong CH lines, were determined by
\citet{Lucatelloetal:2006}. They also discuss the frequency of carbon-enhanced
stars among metal-poor stars, based on the HERES sample.
\citet{Hayeketal:2008} present a detailed abundance analysis of the strongly
r-process enhanced star HE~1219$-$0312 (and an additional such star,
CS~29491-069, from the HK survey), and \citet{HE0338} report on an analysis of
HE~0338$-$3945, a star enriched in r- \emph{and} s-process elements.
Additional HERES papers are in preparation.

A sample of 1777 bright (i.e., $9 < B < 14$) metal-poor candidates selected on
329 HES plates has been presented by \citet{Frebeletal:2006b}. {\FAC}
($B=14.016$; \citealt{Aokietal:2006}) is a member of this group of HES stars,
as is HE~1523$-$0901, a bright ($B=12.186$; \citealt{Beersetal:2007}),
strongly r-process enhanced ($\mathrm{[r/Fe]} = +1.8$) star in which the
\ion{U}{ii} line at $\lambda = 3859.57$\,{\AA} is detected
\citep{Frebeletal:2007b}, which is crucial for determining a
nucleochronometric age of a star. Snapshot spectra of confirmed metal-poor
stars from this sample have been obtained with the AAT/UCLES, Magellan/MIKE,
VLT/UVES, and HET/HRS.

The selection described in this paper includes bright stars, thus many of the
stars already published in \citet{Frebeletal:2006b} are also present in the
sample reported herein. However, the selection criteria for bright stars have
been improved compared to those of Frebel et al.; in particular, we only make
limited use of the HES $B-V$ colors for bright stars, because this color has
been shown by Frebel et al. to be unreliable due to saturation effects for
objects brighter than $B=13$. Instead, we now use a selection criterion based
on $J-K$ colors from the Two Micron All Sky Survey (2MASS;
\citealt{Skrutskieetal:2006}); i.e., a criterion which does not involve any
colors measured from HES spectra. We also extended the selection of bright
stars to 50 HES fields that were not explored by Frebel et al.

\section{Calibrations}\label{Sect:Calibrations}

As explained in Sect. \ref{Sect:HESproperties} above, we wish to make use of
the $B-V$ or $J-K$ color and the strength of the \ion{Ca}{ii}~K line to
perform a quantitative, bias-free selection of metal-poor candidates in the
HES. Over a wide color range, the \ion{Ca}{ii}~K line is strong enough to be
detected even at low metallicities, and at a given color, the KP index
typically varies by more than 2\,{\AA} per dex in [Fe/H] (see
Fig.~\ref{Fig:KPcutoff}), making a selection based on these observables
feasible.  While $J$ and $K$ photometry is available from 2MASS, the $B-V$
color and the KP index can be measured directly from the HES spectra.
However, it is necessary to calibrate these measurements against CCD
photometry and previously obtained medium-resolution spectra, respectively, to
avoid systematic offsets in the HES measurements.

\subsection{$B-V$ color}\label{Sect:BminVHES}

Observations with a Schmidt telescope and an objective prism yield slitless
spectroscopy, from which approximate spectrophotometric information can be
extracted. In the case of the HES, the plate material is sufficiently
homogeneous (as judged from the results presented at the end of this section),
and the wavelength range, covering the full $U$ and $B$ bands, as well as
about half of the $V$ band, is large enough to extract color information
directly from the digital objective-prism spectra. \citet{hespaperIII} defined
so-called spectral half-power points (HPPs) of HES spectra for selecting
quasar candidates by means of their spectral energy distribution.
\citet{HESStarsI} calibrated two of these HPPs onto the $UBV$ photometric
system, yielding estimates of $U-B$ and $B-V$ for all HES sources. The
\citet{HESStarsI} calibration for $B-V$ reached an accuracy of $0.1$\,mag in
the color range $-0.6 < B-V < 2.0$. We slightly improve on that accuracy by
establishing a calibration for $B-V$ which is restricted to the color range
relevant for selecting metal-poor candidates, i.e., $0.3 < B-V < 1.4$.

\begin{figure}[htbp]
  \centering
  \includegraphics[clip=true,bb=64 475 457 771,width=9cm]{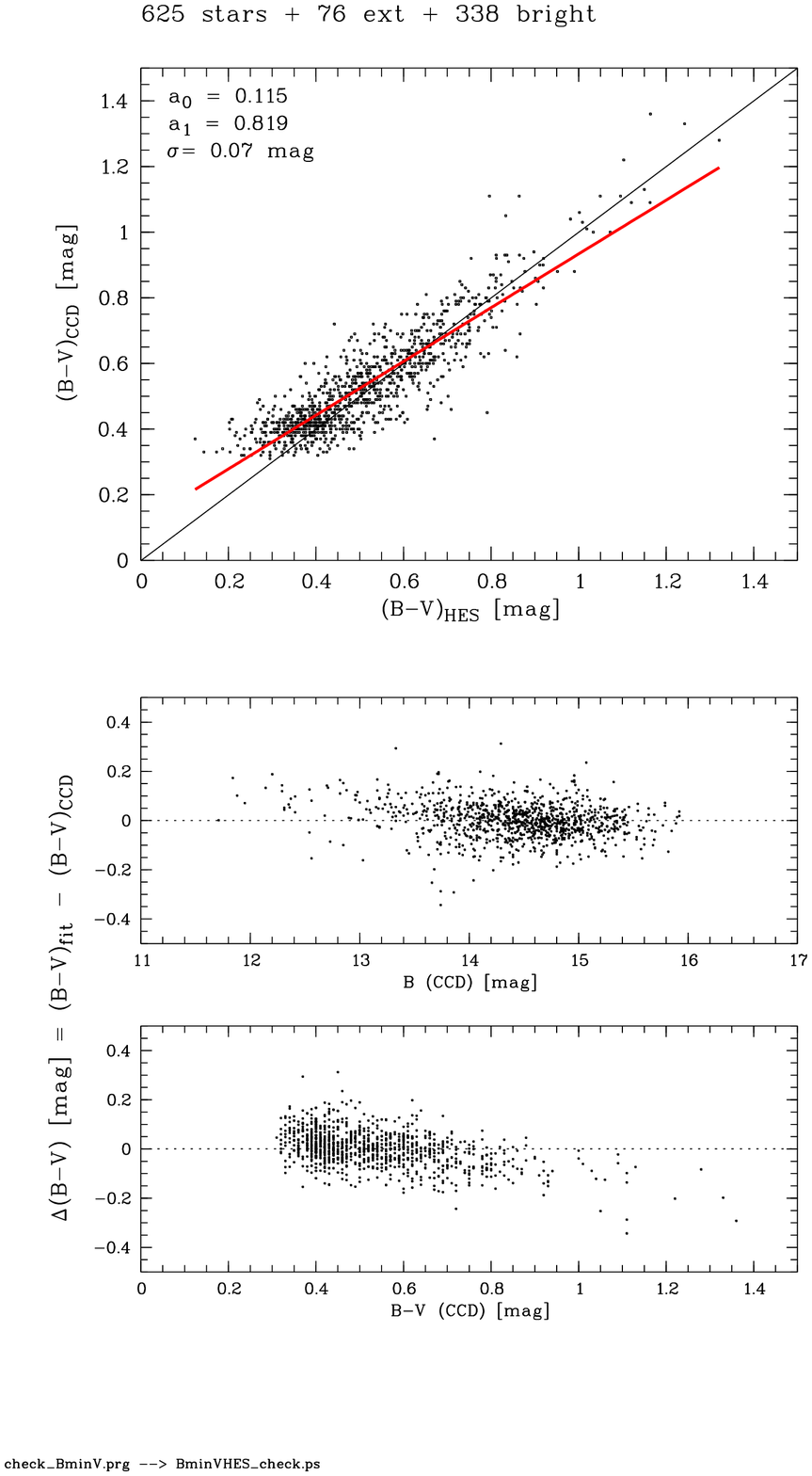}
  \caption{\label{Fig:BminVHES_check} Calibration of $B-V$ colors 
    estimated from HES spectra, using 1039 metal-poor stars from the HK survey
    present on HES plates.}
\end{figure}

To carry out this calibration, we employ 1039 metal-poor stars from the HK
survey in the magnitude range $12 < B < 16$ that are also present on HES
plates, and for which accurate CCD photometry is available. Excluded from the
calibration sample (as well as from the selection of metal-poor stars) were
stars which have been detected in the HES as being influenced by an
overlapping spectrum caused by an object in the dispersion direction
(hereafter referred to as \emph{overlaps}). The calibration sample includes
625 \texttt{stars} (i.e., unsaturated point sources), 338 sources of the
object type \texttt{bright} (i.e., sources above a saturation threshold), and
76 \texttt{ext} objects (i.e., objects that have been classified as extended
sources).

Figure \ref{Fig:BminVHES_check} shows the result of the $B-V$ calibration
exercise for the sample of 1039 HK-survey stars. The 1-$\sigma$ uncertainty is
$0.07$\,mag. This is remarkably low for colors estimated from photographic
plates, especially considering that the calibration involves stars from many
plates, taken over a period of almost 10 years. Calibrating all three classes
of sources separately yielded 1-$\sigma$ uncertainties of $0.06$\,mag
(\texttt{stars}), $0.07$\,mag (\texttt{bright}), and $0.08$\,mag
(\texttt{ext}), respectively. We use these individual calibrations for the
respective source types for selecting metal-poor candidates, whereas
\texttt{ext} sources are excluded from the selection altogether to avoid
galaxies entering the candidate sample as false positives.

\subsection{Line indices}

Line indices compare the continuum level at the center of an absorption line
or feature with its depth relative to the continuum. Indices are a measure of
the integrated absorption of the line or feature, and, analogous to equivalent
widths, they carry units of wavelengths. \citet{Beersetal:1999} defined the KP
index for the \ion{Ca}{ii}~K line, and the GP index for the CH G band. We
calibrated these indices obtained from HES spectra, by comparing with
measurements of these indices in medium-resolution (i.e., $\Delta\lambda \sim
2$\,{\AA}) spectra of more than 2000 HK-survey stars present on HES plates.
The wavelength bands of the indices, as used in the HES, are summarized and
compared to the \citet{Beersetal:1999} definitions in Tab.
\ref{Tab:Passbands}, and illustrated in Fig. \ref{Fig:HESspectrum_indices}.

\begin{table}[htbp]
 \centering
 \caption{Wavelength bands of the KP index for the Ca~K line and the GP index
   for the G band of CH, as employed in the HES (KPHES, GPHES), and compared with the
   definitions of K6, K12, K18 and GP of \citet{Beersetal:1999}.}
 \label{Tab:Passbands}
  \begin{tabular}{lccc}\hline\hline
    Index  & Blue Sideband [{\AA}] & Line Band [\,{\AA}] & Red Sideband [\,{\AA}]\\\hline
    KPHES  &      --        & 3920.0--3946.0 & 4011.0--4065.0\\
    K6     & 3903.0--3923.0 & 3930.7--3936.7 & 4000.0--4020.0\\
    K12    & 3903.0--3923.0 & 3927.7--3939.7 & 4000.0--4020.0\\
    K18    & 3903.0--3923.0 & 3924.7--3942.7 & 4000.0--4020.0\\
    GPHES  & 4246.0--4255.0 & 4281.0--4307.0 & 4446.0--4612.0\\
    GP     & 4247.0--4267.0 & 4297.5--4312.5 & 4362.0--4372.0\\\hline
  \end{tabular}
\end{table}

\subsubsection{The KP index}\label{Sect:KPHES}

Since the resolution of the HES spectra is typically 10\,{\AA} at
\ion{Ca}{ii}~K (depending on the seeing), the spectral lines are broadened to
an extent where essentially no continuum is present between H$_8$ and
\ion{Ca}{ii}~K. Therefore, it was not possible to use the blue sideband for
the KP index defined by \citet{Beersetal:1999}, or any other blue sideband,
for that matter. For the same reason, the red sideband employed by Beers et
al. was replaced by a 54\,{\AA} wide band covering the wavelength region
4011--4065\,{\AA}.

The resolution of the HES spectra does not allow one to distinguish between
the 6\,{\AA}, 12\,{\AA}, and 18\,{\AA} wide line bands for the \ion{Ca}{ii}~K
line defined by \citet{Beersetal:1999}. We therefore do not apply the
band-switching scheme for the line passband suggested by Beers et al. Instead,
a single 26\,{\AA} wide band centered on \ion{Ca}{ii}~K was used. This band is
sampled by only 4 pixels in the HES spectra.

\begin{figure}[htbp]
  \centering
  \includegraphics[clip=true,bb=70 114 461 769,width=9cm]{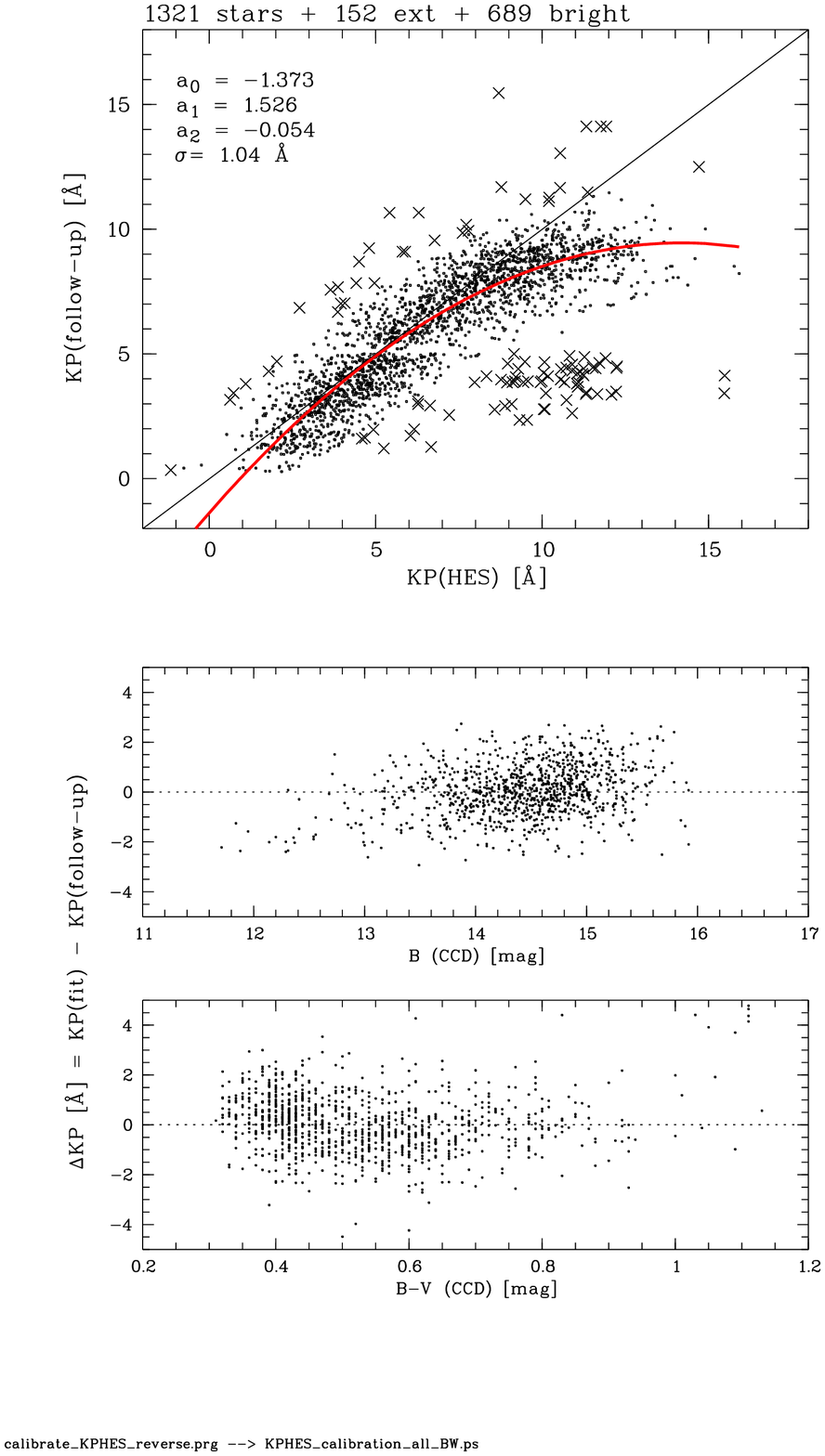}
  \caption{\label{Fig:KPHES_calibration} Calibration of the KP index for the
    \ion{Ca}{ii}~K line, using 2162 metal-poor stars from the HK survey
    present on HES plates. Stars that have been rejected from the fit as
    outliers are shown as crosses. As can be seen in the uppermost panel, the
    use of a broader line passband appears to lead to a later onset of
    saturation of the KP index compared to the follow-up spectra.
}
\end{figure}

We proceed by fitting a second-order polynomial to the data points from the
HES KP measurements (hereafter KPHES) and those from the medium-resolution
follow-up spectra (hereafter KPfu), after rejecting 104 stars as $> 2\,\sigma$
outliers in a first iteration of the fit (see Fig.
\ref{Fig:KPHES_calibration}), leaving 2162 stars. In Fig.
\ref{Fig:KPHES_calibration} one can see a distinct group of 53 stars (i.e.,
$\sim 2\,\%$ of the sample of calibration stars) with KPHES in the range
8--15\,{\AA} and $\mathrm{KPfu}<5$\,{\AA}. Inspection of the HES spectra
reveals that these are spectra of cool stars affected by strong molecular
features bluewards of the Ca~K line. As has been shown by
\citet{Cohenetal:2005}, the presence of CH and CN lines in the continuum bands
used for the measurement of the KP index leads to systematically too low KP
index values. In Fig.~2 of \citet{Cohenetal:2005} it can be seen that the blue
continuum band is more strongly affected by the CH and CN lines than is the
red continuum band, and since the blue continuum band is not used in the HES,
a systematic offset between KPHES and KPfu results.

The 1-$\sigma$ scatter of KPHES around
the adopted calibration relation is 1.0\,{\AA}. Reading from panel \emph{a} of
Fig. \ref{Fig:KPcutoff}, this translates into uncertainties in [Fe/H] of
about 1.0/0.5/0.2/0.3\,dex at $(B-V)_0 = 0.3$/0.4/0.7/1.0, respectively. The
uncertainties in [Fe/H] due to errors in $(B-V)_0$ are negligible compared to
those caused by the errors in KP. For faint or very bright HES stars, the
uncertainties in [Fe/H] are higher by up to a factor of about 2, since the
errors in KP are larger than 1.0\,{\AA} (see Fig.~\ref{Fig:sigKP_BHES}).
However, the numbers listed above indicate that for cool giants in the
magnitude range $B\sim 13$--$16$, [Fe/H] can be estimated from HES spectra
with an accuracy that is only about a factor of two worse than estimates based
on moderate-resolution follow-up spectra.


The use of a single continuum passband might lead to a correlation of KP with
the colors of the stars, because the varying spectral energy distributions
lead to a change in continuum slope, altering the line depth relative to a
pseudo-continuum defined by the continuum band on the red side of the line
only. However, in a plot of the residual of the KP measurements against $B-V$,
we do not see any strong systematic effects of this sort (see lower panel of
Fig.~\ref{Fig:KPHES_calibration}). This is probably due to the fact that the
selection of metal-poor candidates is restricted to a relatively narrow
effective temperature range of approximately 4000-6600\,K.

\subsubsection{The GP index}

For the measurement of the GP index from the HES spectra, we usd continuum
passbands covering 4246--4255\,{\AA} and 4446--4612\,{\AA}, and a 26\,{\AA}
wide band centered on the bandhead of the G band (see Fig.
\ref{Fig:HESspectrum_indices}).

\begin{figure}[htbp]
  \centering
  \includegraphics[clip=true,bb=70 477 461 769,width=9cm]{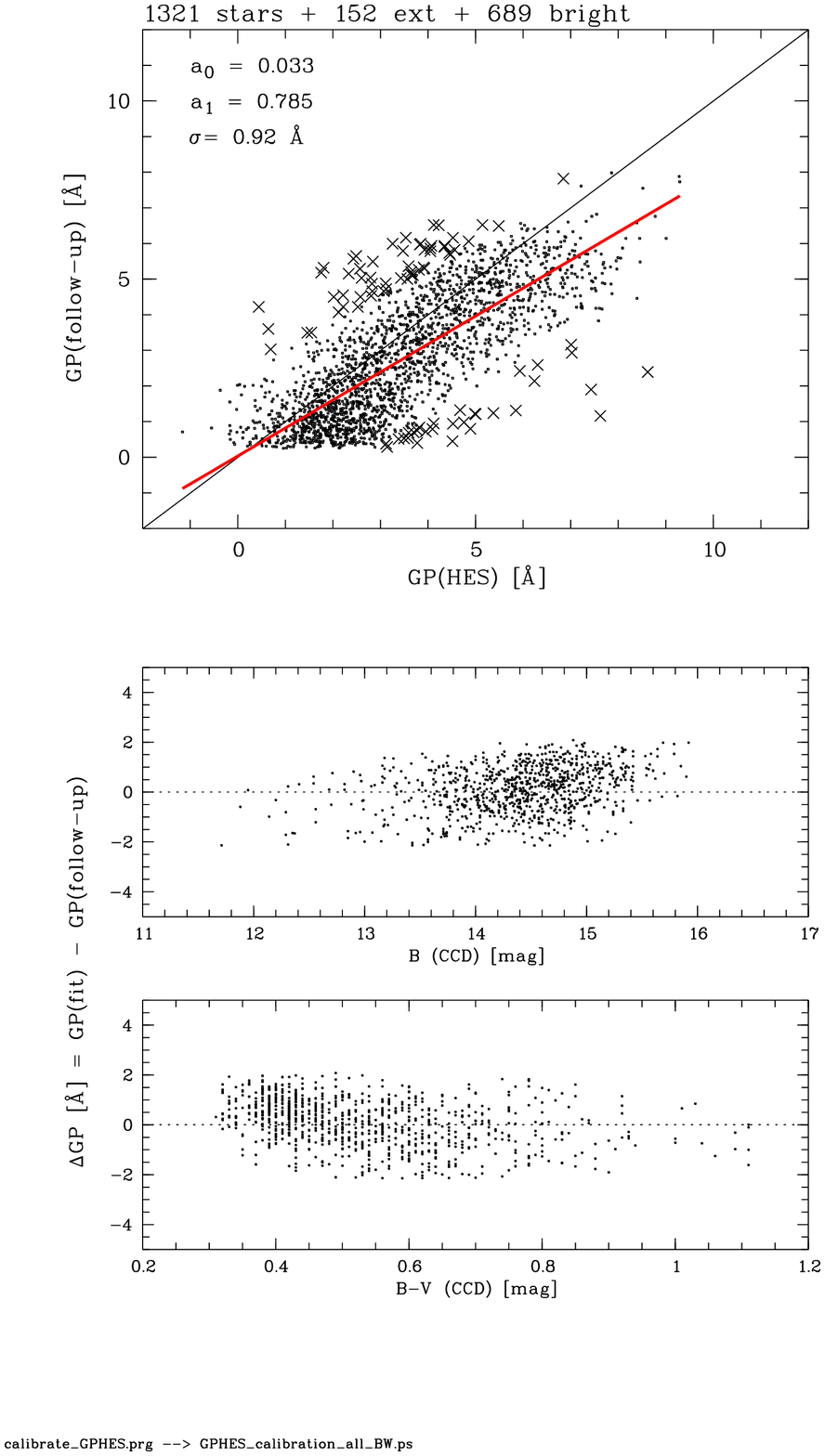}
  \caption{\label{Fig:GPHES_calibration} Calibration of the GP index for the
    CH G band, using 2162 metal-poor stars from the HK survey present on HES 
    plates. Stars that have been rejected from the fit as outliers are shown
    as crosses.}
\end{figure}

The adopted calibration relation is a linear fit to 1990 data points.
Measurements for 86 stars have been rejected as $> 2\,\sigma$ outliers. The
1-$\sigma$ scatter of the GP measurements from the HES spectra around the fit is
0.9\,{\AA}, which would result in an uncertainty of $\sim 0.2$\,dex in [C/Fe]
if other error sources are neglected (in particular, the errors in KP and and
color).

We emphasize that we \emph{do not} involve GP in the selection of metal-poor
candidates, but the index is useful for identifying candidate carbon-enhanced
stars among the metal-poor candidates, if desired. For reference we list in
Tab.~A.1 estimates of [C/Fe] obtained for all HES metal-poor candidates,
derived using the methods of \citet{Rossietal:2005}. These estimates are
based on measurements of the GP and KP indices from the HES spectra.

\begin{figure*}[tp]
  \centering
  \includegraphics[clip=true,bb=49 354 526 684,width=\textwidth]{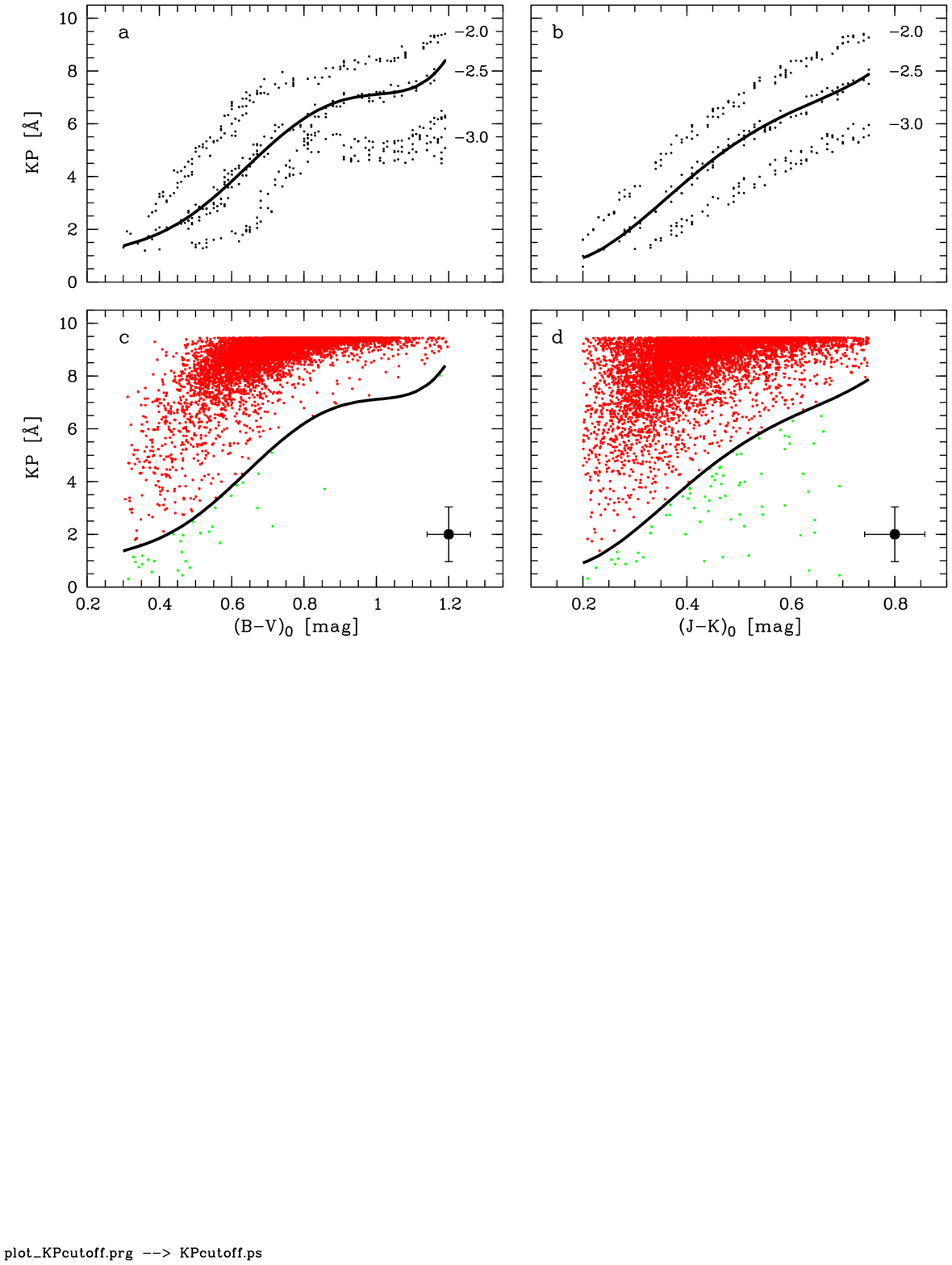}
  \caption{\label{Fig:KPcutoff} Determination of cutoff lines for selecting
    metal-poor candidates (panels \emph{a} and \emph{b}), and their application
    to all stars on HES plate \#12073, ESO field 147 (panels \emph{c} and
    \emph{d}). In panels \emph{a} and \emph{b}, simulated stars with [Fe/H]
    within 0.1\,dex of the indicated metallicities are shown. To the data
    points corresponding to stars of $\mathrm{[Fe/H]}=-2.5\pm 0.1$, a
    polynomial was fitted (thick black lines). In the lower panels, typical
    error bars of the measurements of the KP index and $(B-V)$ color from the
    HES spectra, as well as the typical 1-$\sigma$ uncertainty of the 2MASS
    $(J-K)$ color, are shown. The larger uncertainties of the $(J-K)$ colors
    lead to more candidates below the selection cutoff compared to the
    selection in the KP/$(B-V)_0$ parameter space. The sharp cut at
    $\mathrm{KP}\sim 9.5$\,{\AA} is due to the fact that the KP index cannot
    exceed this value due to the calibration of that index described in Sect.
    \ref{Sect:KPHES}. As can be seen in the uppermost panel of Fig.
    \ref{Fig:KPHES_calibration}, the fitted calibration curve reaches a
    maximum at this value. }
\end{figure*}

\section{Selection of candidate metal-poor stars}\label{Sect:Selection}

In this section the procedures for selecting metal-poor candidates in the HES
are outlined. We aim to develop a selection that (a) does not introduce an
effective-temperature related bias, and that (b) restricts the selected stars
to a manageable number of candidates for conducting medium-resolution
spectroscopic follow-up observations, while maintaining a reasonable level of
completeness. Since we are mainly interested in the most metal-poor stars, we
target a completeness as high as possible for stars with $\mathrm{[Fe/H]} <
-3.5$, while some incompleteness at $\mathrm{[Fe/H]}\sim -3.0$ can be
tolerated, and a rejection of as many stars with $\mathrm{[Fe/H]}> -2.5$ as
possible is desired.

\subsection{Selection criteria}\label{Sect:Criteria}

\subsubsection{The KP/$(B-V)_0$ selection}\label{Sect:KPversBV0}

The first of our selection criteria employs a cutoff line in the parameter
space KP versus $(B-V)_0$. The measurements of $B-V$ and KP from HES spectra
have been discussed in Sections \ref{Sect:BminVHES} and \ref{Sect:KPHES} above,
respectively.

The reddening has been determined using the maps of \citet{Schlegeletal:1998}.
By comparison of measured (de-reddened) $B-V$ colors with predictions of this
color from the HP2 index for H$\delta$, measured in moderate-resolution
spectra, \citet{Beersetal:2002} found that at $E(B-V) > 0.10$ the reddening of
the Schlegel et al. maps is too high. That is, $(B-V)_0$, as computed from CCD
$B-V$ photometry and $E(B-V)$ from the Schlegel et al. maps is systematically
too blue for stars with $E(B-V) > 0.10$ compared to the predictions of
$(B-V)_0$ from HP2. This problem was previously recognized by
\citet{Arce/Goodman:1999}. To rectify it, \citet{Beersetal:2002} reduced any
reddening in excess of $E(B-V) = 0.10$ by 35\,\%. We adopt the same procedure.

The KP cutoff line was defined in the following way. An artificial sample of
10,000 stars was created, by computing with a random number generator values
of $(B-V)_0$ equally distributed in the range $0.3 \le (B-V)_0 \le 1.2$, and
values of KP equally distributed in the range $0.0 < \mathrm{KP} < 12$. These
pairs of values were then converted into [Fe/H], using an improved version of
the method of \citet{Beersetal:1999}, which involves more calibration stars
(Beers et al., in preparation). All data points yielding
$\mathrm{[Fe/H]}=-2.5\pm 0.1$ were selected, and a polynomial of 5th order was
fitted to these data points. The weight of the points at $(B-V)_0 < 0.45$ and
$(B-V)_0 > 1.00$ was increased by a factor of 10 to improve the quality of the
fit. The result is displayed in panel \emph{a} of Fig.  \ref{Fig:KPcutoff}.
The cutoff line nicely runs parallel to the lower edge in the KP distribution
at a given color (i.e., the marginal distribution in KP). That is, the lines
could alternatively have been determined empirically, using a cutoff-location
algorithm as was employed by \citet{hespaperIII} for selecting UV-excess
objects in the HES.

An object is selected as a metal-poor candidate if it falls into the color
range $0.3 \le (B-V)_0 \le 1.2$ (roughly corresponding to $4000\,\mathrm{K} <
\teffm < 6600\,\mathrm{K}$), and if its KP value is below the cutoff for the
given $(B-V)_0$, but above a threshold for a significant detection of an
emission line (see Sect.  \ref{Sect:sigKP} for details).  Further ingredients
for the KP/$(B-V)_0$ selection are the average $S/N$ per pixel in the $B_J$
band and in the \ion{Ca}{ii}~H and K region. This aims at rejecting spectra
that are too noisy for a sensible candidate selection. A candidate is rejected
if (a) the average $S/N$ per pixel in the $B_J$ band is lower than 5/1, or if
(b) the average $S/N$ in the \ion{Ca}{ii}~H and K region is lower than 5/1.
The method for estimating the noise per pixel of the HES spectra has been
described by \citet{HESStarsI}. 

Criterion (a) typically corresponds to a magnitude limit of $B_J < 17.5$,
while criterion (b) results in a restriction to objects that are about 1\,mag
brighter in the $B_J$ band.  However, the magnitude limit varies considerably
from plate to plate, depending critically on the conditions under which the
plates were exposed (i.e., sky background level and seeing). Bright stars with
(partly) saturated HES spectra can have a higher $S/N$ at \ion{Ca}{ii}~H and K
compared to the average $S/N$ in the $B_J$ band, so that the $S/N$ criterion
(a) rejects the brightest (i.e., $B < 11$--$13$, depending on plate quality)
stars, while at the faint end only criterion (b) has an effect.

Finally, all candidates with spectra affected by an overlap, according to the
automated overlap detection algorithm, are rejected.

\subsubsection{The KP/$(J-K)_0$ selection}\label{Sect:KPversJK0}

A selection cutoff in the KP versus $(J-K)_0$ parameter space has been
determined analogous to the cutoff in the KP versus $(B-V)_0$ parameter space,
except that a polynomial of 4th order was fit, and equal weights were assigned
to all data points. The result is shown in panel \emph{b} of Fig.
\ref{Fig:KPcutoff}.

The $J$ and $K$ magnitudes for the HES stars were retrieved from the 2MASS
All-Sky Data Release. The cross-correlation of the two catalogs was
accomplished by adopting the closest 2MASS point source within a search box of
$5\arcsec \times 5\arcsec$ as the matching object. We note that the 1-$\sigma$
uncertainty of the HES coordinates is the same as the uncertainty of the DSS-I
plate solutions, which is smaller than $1\arcsec$ for point sources (see,
e.g., \citealt{Veron-Cetty/Veron:1996}). The 2MASS astrometry is even more
accurate; i.e., $\sigma = 0.1\arcsec$ for sources brighter than $K_S = 14$
\citep{Skrutskieetal:2006}. We adopted a search box considerably larger than
the combined uncertainties to be able to identify the corresponding 2MASS
source even in case of HES objects suffering from 5-$\sigma$ astrometric
errors. Over 99\,\% of the HES sources fulfilling the $S/N$ criteria (a) and
(b) listed above were successfully identified in 2MASS.

We only make use of $J$ and $K$ magnitudes which were labeled with the
photometric quality flags ``A'', ``B'', or ``C'', and which had their blend and
contamination flags set to 1 (i.e., one component present only), and 0 (i.e.,
source not affected by any artifacts), respectively. Applying these quality
criteria results in the rejection of the $J$ and $K$ magnitudes for about 1\,\%
of the HES sources identified in 2MASS.

The $J-K$ colors were de-reddened by adopting the relation $E(J-K) =
0.535\cdot E(B-V)$, derived from Table 6 of \citet{Schlegeletal:1998}. Here,
$E(B-V)$ values in excess of $0.10$ have been scaled down in the same manner
as was done for the de-reddening of $B-V$ (see Sect. \ref{Sect:KPversBV0}).

We restrict the candidate selection to stars in the color range $0.2 <
(J-K)_0 < 0.75$. The application of the KP cutoff, the $S/N$ criteria, and the
rejection of overlaps follows the same procedures as the KP/$(B-V)_0$ selection
described in the previous section.

\subsubsection{Non-detection of the \ion{Ca}{ii}~K line}\label{Sect:sigKP}

For stars near the main-sequence turnoff, the KP cutoff lines shown in Fig.
\ref{Fig:KPcutoff} approach levels comparable to the measurement uncertainty
of KP. Therefore, even if a star would have no \ion{Ca}{ii}~K line, it might
be rejected because random noise might result in the measurement of a KP index
above the cutoff for turnoff stars. Thus it is mandatory to relax the
selection criteria for these stars. This is achieved by including stars into
the candidate sample if their KP index is above the cutoff, but a
\ion{Ca}{ii}~K line is not significantly detected in the HES spectrum.

\begin{figure}[htbp]
  \centering
  \includegraphics[clip=true,bb=64 476 458 774,width=9cm]{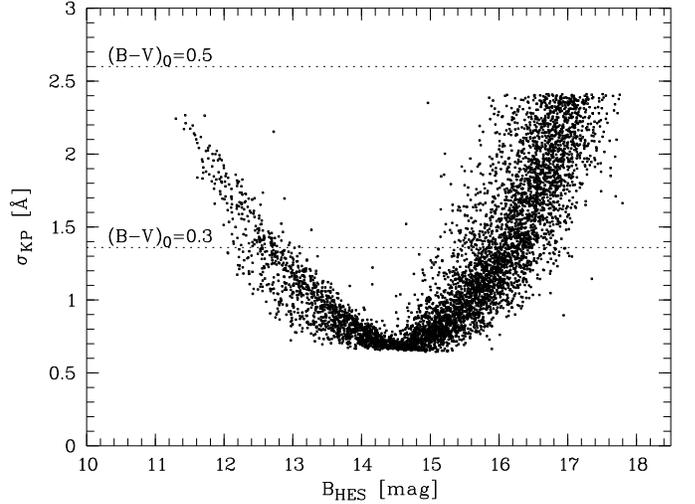}
  \caption{\label{Fig:sigKP_BHES} The individual 1-$\sigma$ detection limit of 
    \ion{Ca}{ii}~K in HES spectra on HES plate \#1528 (ESO field 894) as
    a function of the HES $B$ magnitude. Overplotted are the selection cutoff
    lines for stars of $(B-V)_0=0.3$ and $0.5$ (dashed lines). The position of
    the lower line demonstrates that very bright or faint stars with
    $(B-V)_0=0.3$ are rejected by the KP/$(B-V)_0$ criterion even if a
    \ion{Ca}{ii}~K line is not significantly detected (i.e.,
    $\sigma_{\mathrm{KP}}$ is larger than the cutoff value of KP in these
    cases); hence the selection needs to be relaxed in these cases, which is
    achieved with the \ion{Ca}{ii}~K non-detection criterion (see Sect.
    \ref{Sect:sigKP}). The position of the upper line shows that for stars
    $(B-V)_0\gtrsim 0.5$, the selection does not have to be relaxed, because
    $\sigma_{\mathrm{KP}}$ is smaller than the cutoff value for stars in the
    whole magnitude range, and hence the \ion{Ca}{ii}~K line is significantly
    detected in these cases. }
\end{figure}

Figure \ref{Fig:sigKP_BHES} shows the 1-$\sigma$ detection limit of
\ion{Ca}{ii}~K for all overlap-free spectra belonging to the source types
\texttt{stars} or \texttt{bright} on one randomly selected HES plate. As can
be seen by comparing this figure with Fig. \ref{Fig:KPcutoff}, the selection
criterion is considerably relaxed for faint or very bright turnoff stars
compared to the KP cutoffs. However, for stars at $(B-V)_0 \ga 0.5$ or
$(J-K)_0 \ga 0.3$, the additional criterion does not have any effect, because
the KP cutoff for the stars in this color range is higher than the 1-$\sigma$
detection limit of \ion{Ca}{ii}~K at any magnitude; a star having a KP value
corresponding to a 1-$\sigma$ detection of \ion{Ca}{ii}~K would be selected
anyway.

To understand the behavior of the bright end of the curve shown in Fig.
\ref{Fig:sigKP_BHES}, one has to consider that the HES objective-prism plates
are photographic plates that have been digitized with a microdensitometer. As
one approaches the brightest magnitudes, and correspondingly high photographic
densities of the objective-prism spectra, significantly lower numbers of
photons pass through the photographic plate to be recorded by the scanner,
thereby decreasing the $S/N$ of the brighter HES spectra.

A star is selected if it falls into the color range $0.2 < (J-K)_0 < 0.75$ or
$0.3 \le (B-V)_0 \le 1.2$, and if \ion{Ca}{ii}~K is not significantly detected
in either absorption or emission. The last part of the criterion aims at
rejecting dMe stars or other objects having \ion{Ca}{ii}~H and K in emission.
In addition, the same $S/N$ criteria as in the KP/$(B-V)_0$ and KP/$(J-K)_0$
are applied, and spectra affected by overlaps are rejected.

\subsubsection{Application to the HES}

Below we summarize the selection applied to each of the three HES source
types.
\begin{description}
\item[\texttt{stars} (unsaturated point sources):] An object is selected as a
  metal-poor candidate if it fulfills one or more of the criteria described in
  Sections \ref{Sect:KPversBV0}--\ref{Sect:sigKP} above; i.e., the
  KP/$(B-V)_0$, KP/$(J-K)_0$, or the \ion{Ca}{ii}~K non-detection criterion.
\item[\texttt{bright} (bright sources above a saturation threshold):] An
  object is selected as a metal-poor candidate if it either fulfills the
  KP/$(J-K)_0$ criterion (Sect. \ref{Sect:KPversJK0}) or the \ion{Ca}{ii}~K
  detection criterion (Sect. \ref{Sect:sigKP}), or both. We do not apply the
  KP/$(B-V)_0$ criterion for the bright stars, because it has been shown by
  \citet{Frebeletal:2006b} that the HES $B-V$ colors are not reliable for the
  brightest stars due to saturation effects. Note, however, that the
  \ion{Ca}{ii}~K non-detection criterion does involve the HES $B-V$ color, but
  only for restricting the color range of the candidates to be selected.
\item[\texttt{ext} (extended sources):] All sources classified as extended are
  rejected to avoid contamination of the candidate sample with galaxies.
\end{description}

The application of the KP/$(B-V)_0$ and KP/$(J-K)_0$ criteria to all stars on
one HES plate is illustrated in panels \emph{c} and \emph{d} of Fig.
\ref{Fig:KPcutoff}, respectively.

\begin{figure*}[htbp]
  \centering
  \includegraphics[clip=true,bb=65 508 372 685,width=9cm]{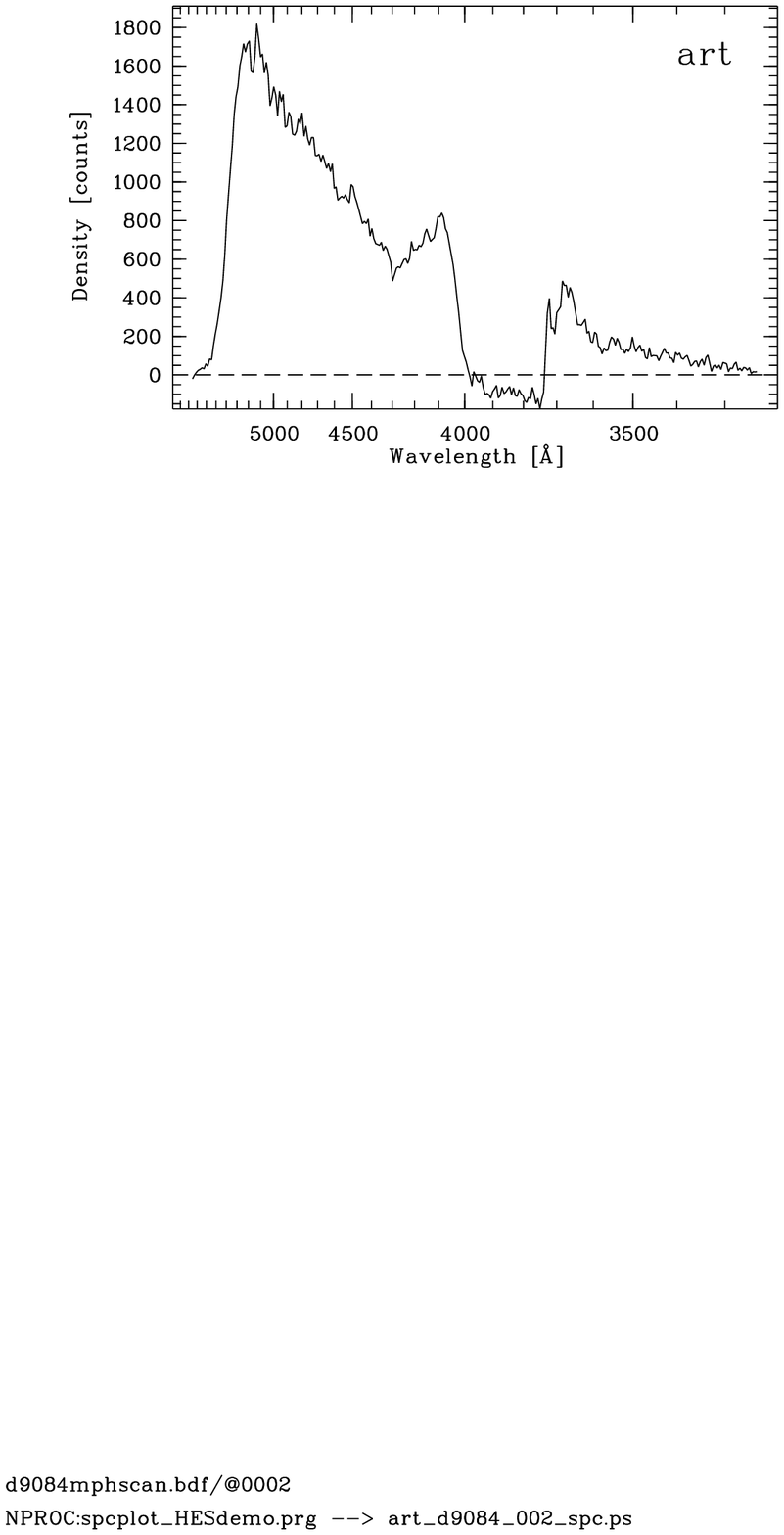}\hspace{0.5ex}
  \includegraphics[clip=true,bb=65 508 372 685,width=9cm]{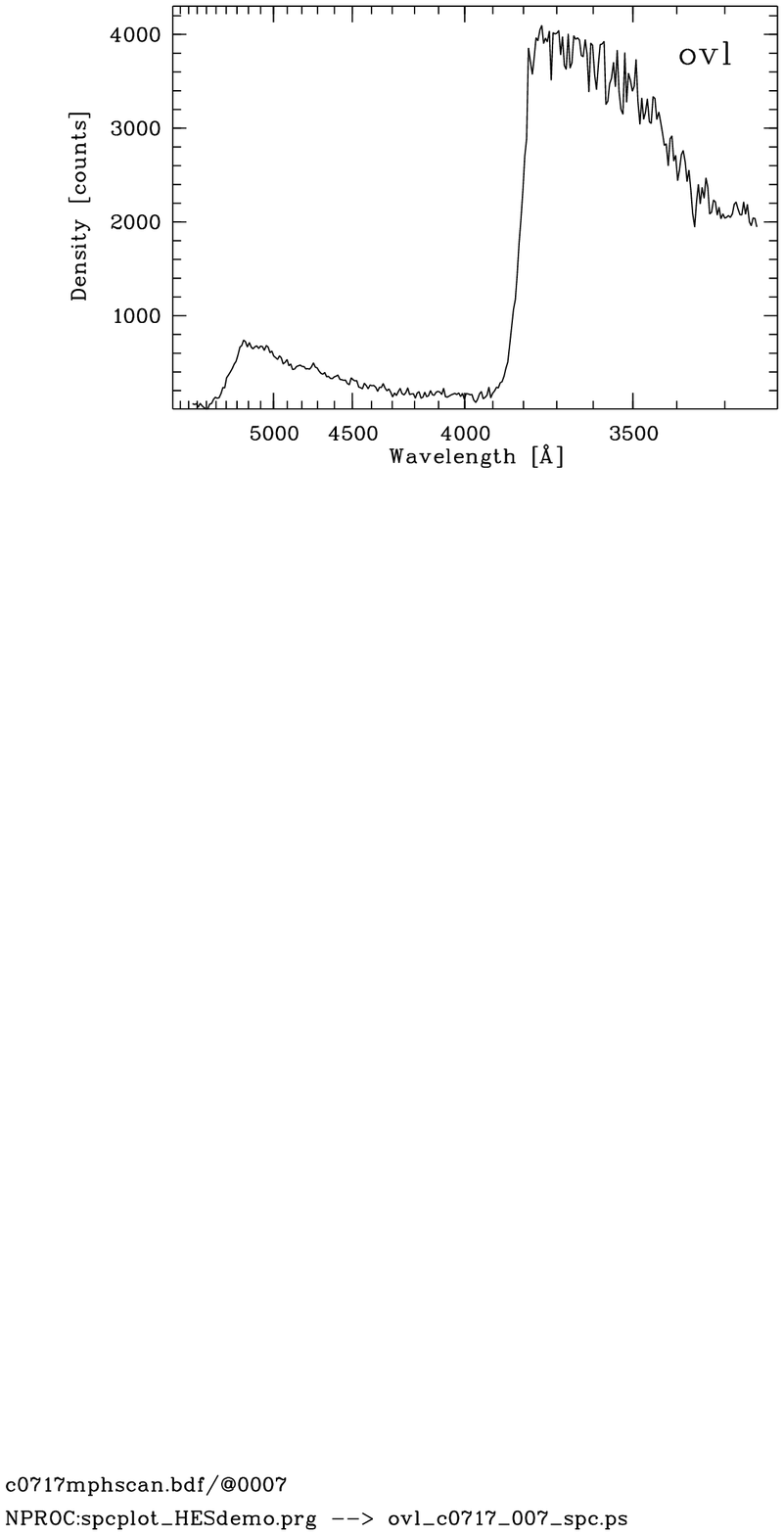}\\[1ex]
  \includegraphics[clip=true,bb=65 508 372 685,width=9cm]{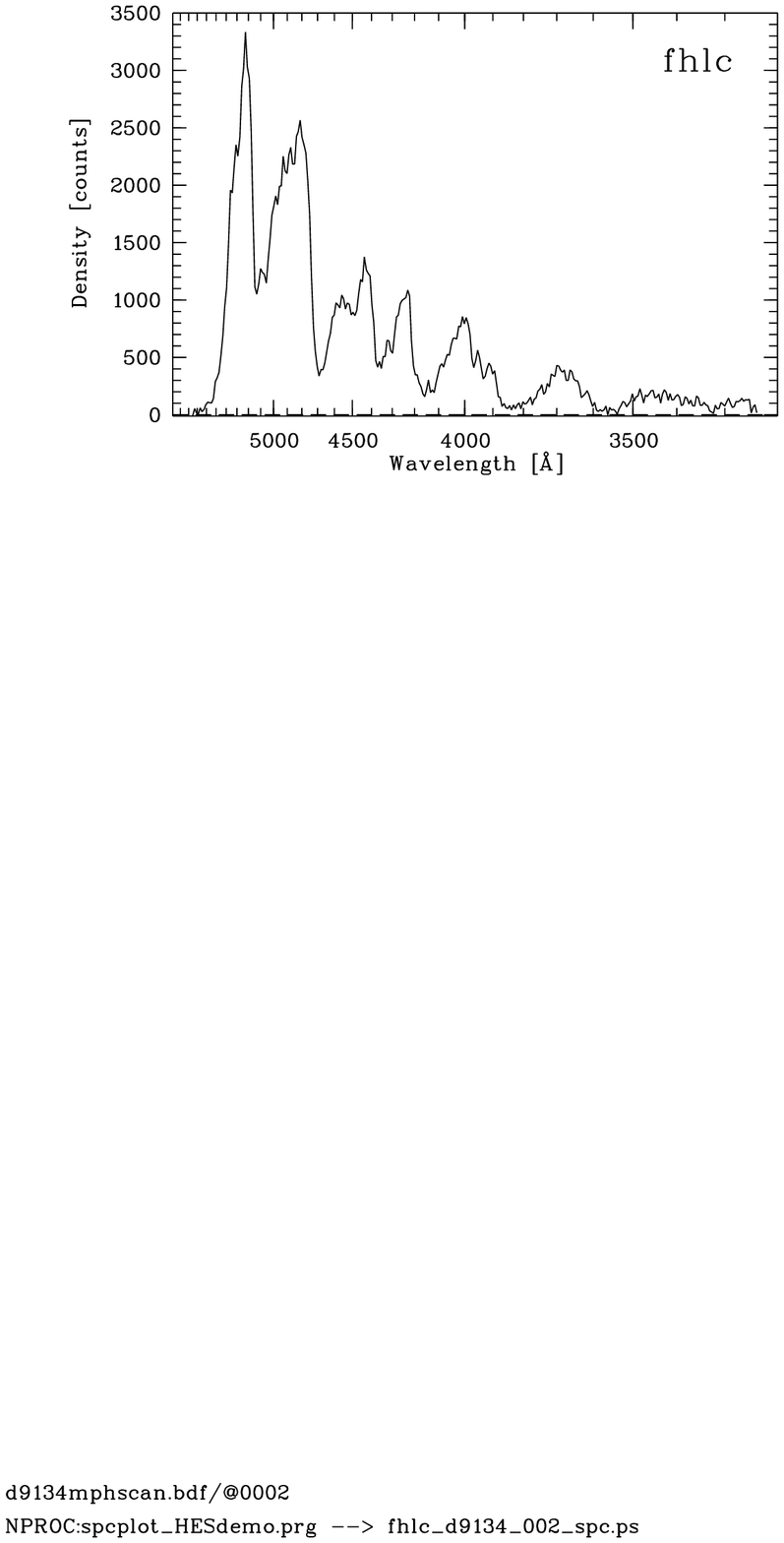}\hspace{0.5ex}
  \includegraphics[clip=true,bb=65 508 372 685,width=9cm]{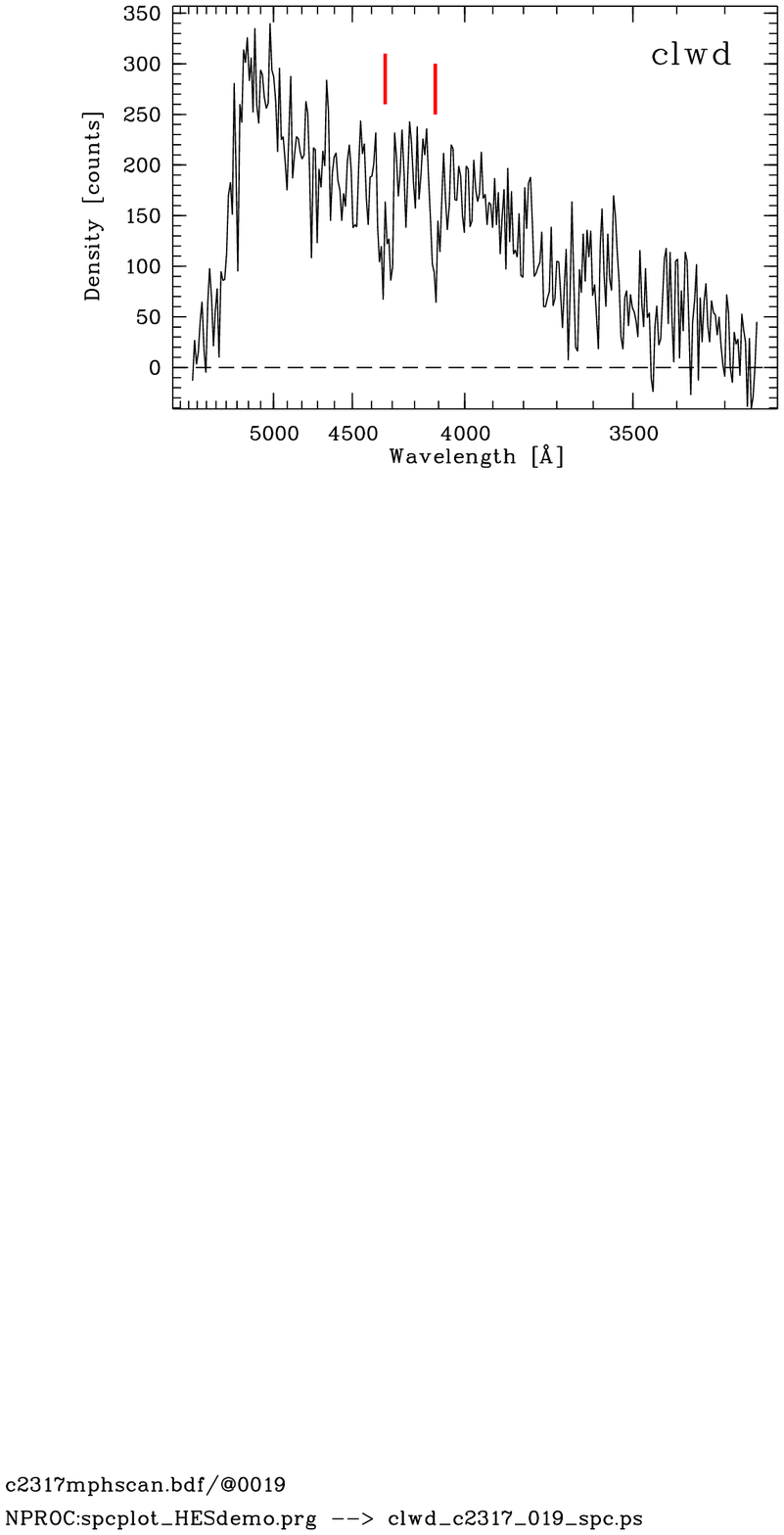}\\[1ex]
  \includegraphics[clip=true,bb=65 484 372 685,width=9cm]{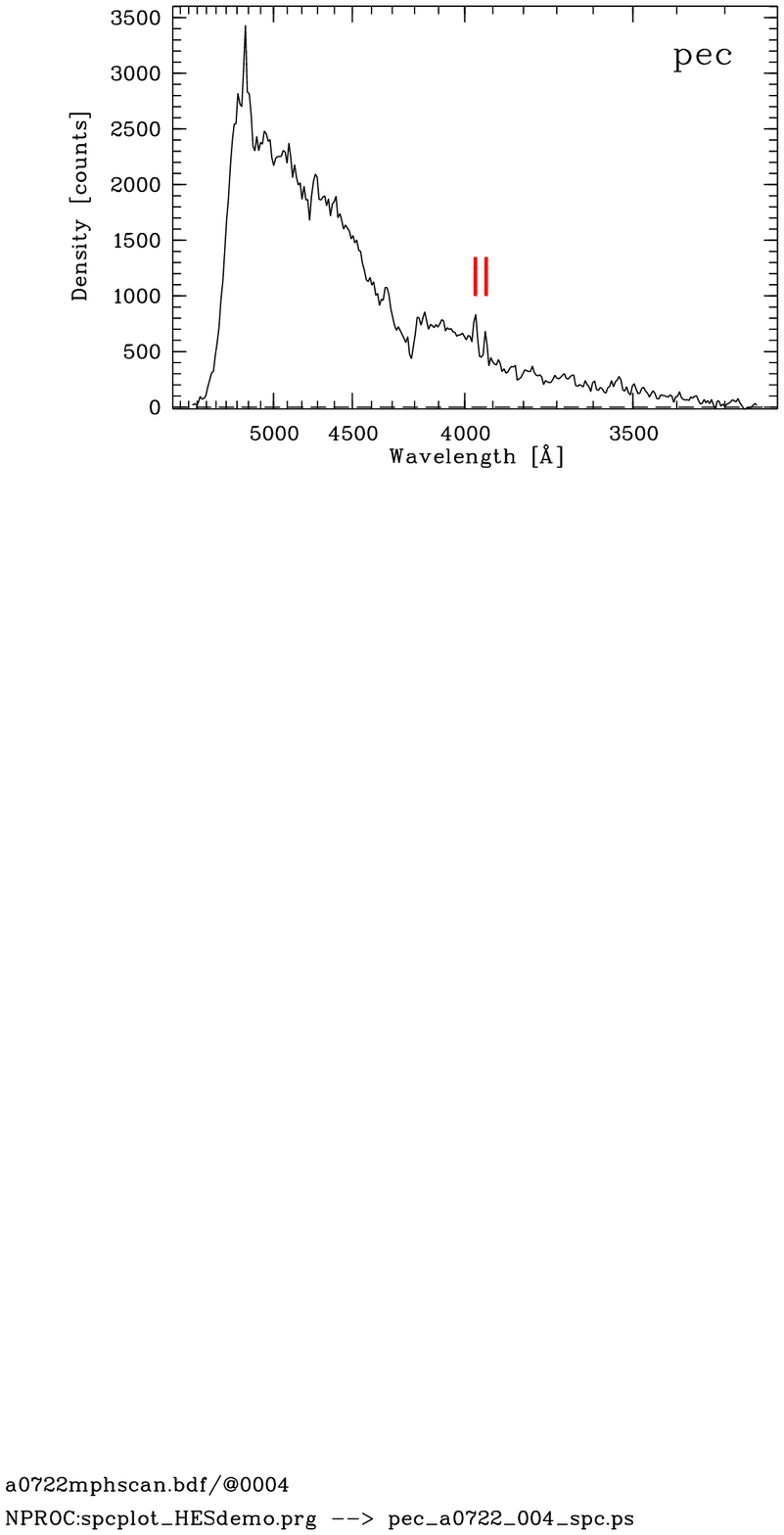}\hspace{0.5ex}
  \includegraphics[clip=true,bb=65 484 372 685,width=9cm]{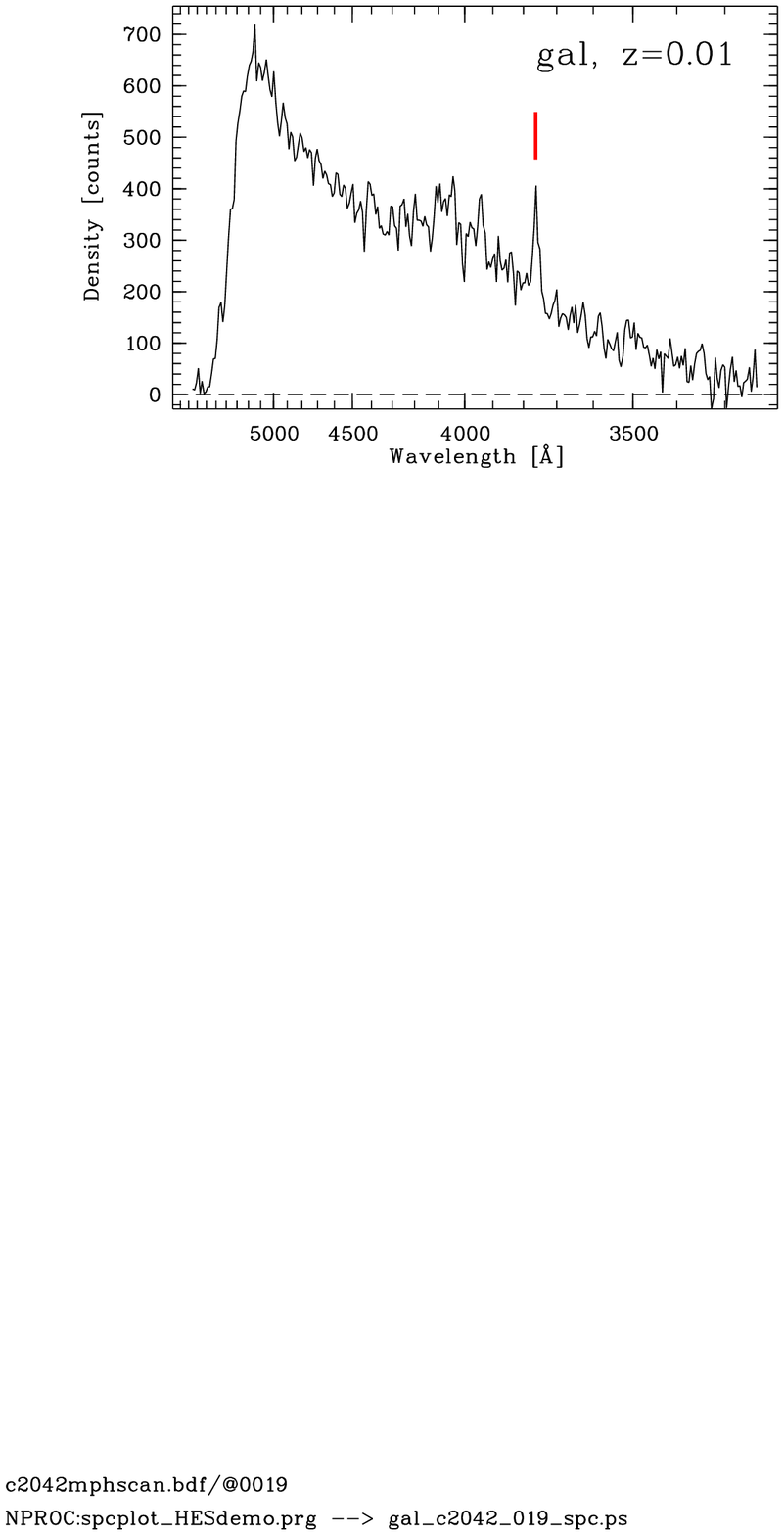}
  \caption{\label{Fig:reject_spectra} HES example spectra of rejected
  metal-poor candidates. \texttt{art} -- spectrum corrupted due to plate artifact,
  in this case by a scratch in the photographic emulsion; \texttt{ovl} --
  spectrum disturbed by bright object in dispersion direction, causing
  overlapping spectra; \texttt{fhlc} -- faint high latitude carbon star;
  \texttt{clwd} -- cool white dwarf, identifiable by the very broad lines of
  H$\gamma$ and H$\delta$ as well as the weak Balmer jump; \texttt{pec} --
  peculiar star (\ion{Ca}{ii}~H and K in emission); \texttt{gal} -- galaxy, showing
  [OII] 3727\,{\AA} in emission and at a redshift of $z=0.01$. }
\end{figure*}

\begin{figure}[htbp]
  \centering
  \includegraphics[clip=true,bb=52 86 442 779,width=9cm]{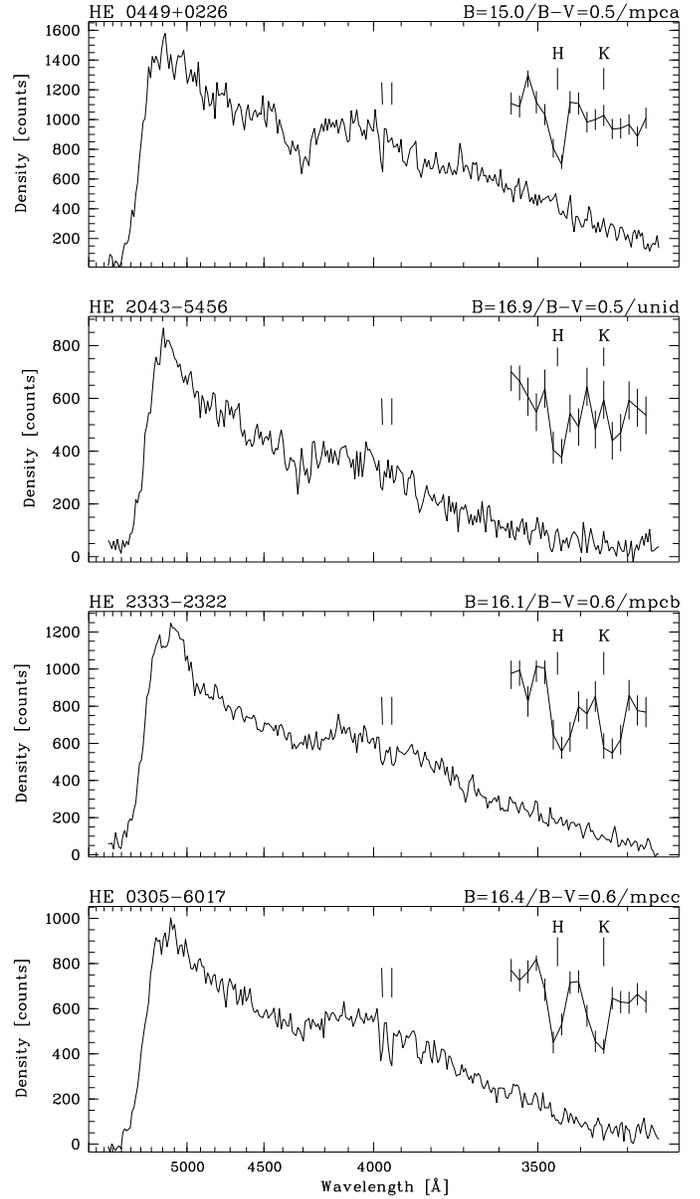}
  \caption{\label{Fig:HESmphs_examples} HES example spectra of candidates of
    the classes \texttt{mpca}, \texttt{unid}, \texttt{mpcb}, \texttt{mpcc}. In
    the upper right corner of each panel, an enlargement of the \ion{Ca}{ii}~H
    and K wavelength region is shown (note that \ion{Ca}{ii}~H is blended with
    H$\epsilon$).}
\end{figure}

\subsection{Visual inspection}\label{Sect:Inspection}

The candidates selected by the criteria described above have to be visually
inspected to reject false positives caused, e.g., by emulsion flaws,
scratches, dust, or other artifacts on the objective-prism plates, or by
overlapping spectra that were not recognized by our automated procedures. We
inspected the extracted HES spectrum together with the raw spectrum on the
digitized objective-prism plate, as well as the corresponding regions on the
relevant DSS-I direct plate.

To avoid any subjective selection biases during the visual inspection, only
candidates were rejected that are clearly disturbed by artifacts or
overlapping spectra, or clearly identifiable as ``peculiar objects'' (e.g.,
stars exhibiting \ion{Ca}{ii}~H and K in emission, or other emission-line
objects), galaxies, cool white dwarfs, etc. In Fig.  \ref{Fig:reject_spectra}
we show example spectra of such objects.

We also applied a subjective ranking of the candidates, by assigning them to
one of the following classes: \texttt{mpca} -- clearly no \ion{Ca}{ii}~K line
detected; \texttt{unid} -- it is unclear whether a \ion{Ca}{ii}~K line is
detected; \texttt{mpcb} -- a weak \ion{Ca}{ii}~K line is detected;
\texttt{mpcc} -- a strong \ion{Ca}{ii}~K line is detected. HES spectra of
candidates classified in this way are shown in
Fig.~\ref{Fig:HESmphs_examples}. This ranking is made for the specific purpose
of prioritizing the follow-up spectroscopy target lists, allowing us to
optimize the observational strategy. We emphasize that for statistical
studies, such as determining the metallicity distribution function (MDF) of
the Galactic halo, \emph{complete} samples have to be compiled; i.e.,
\emph{all} candidates selected on a given set of plates have to be followed up
spectroscopically, regardless of their candidate class. Otherwise,
non-quantifiable selection biases may result.

\section{Evaluation of the selection}\label{Sect:Evaluation}

\begin{figure}[htbp]
  \centering
  \includegraphics[clip=true,bb=63 220 457 770,width=9cm]{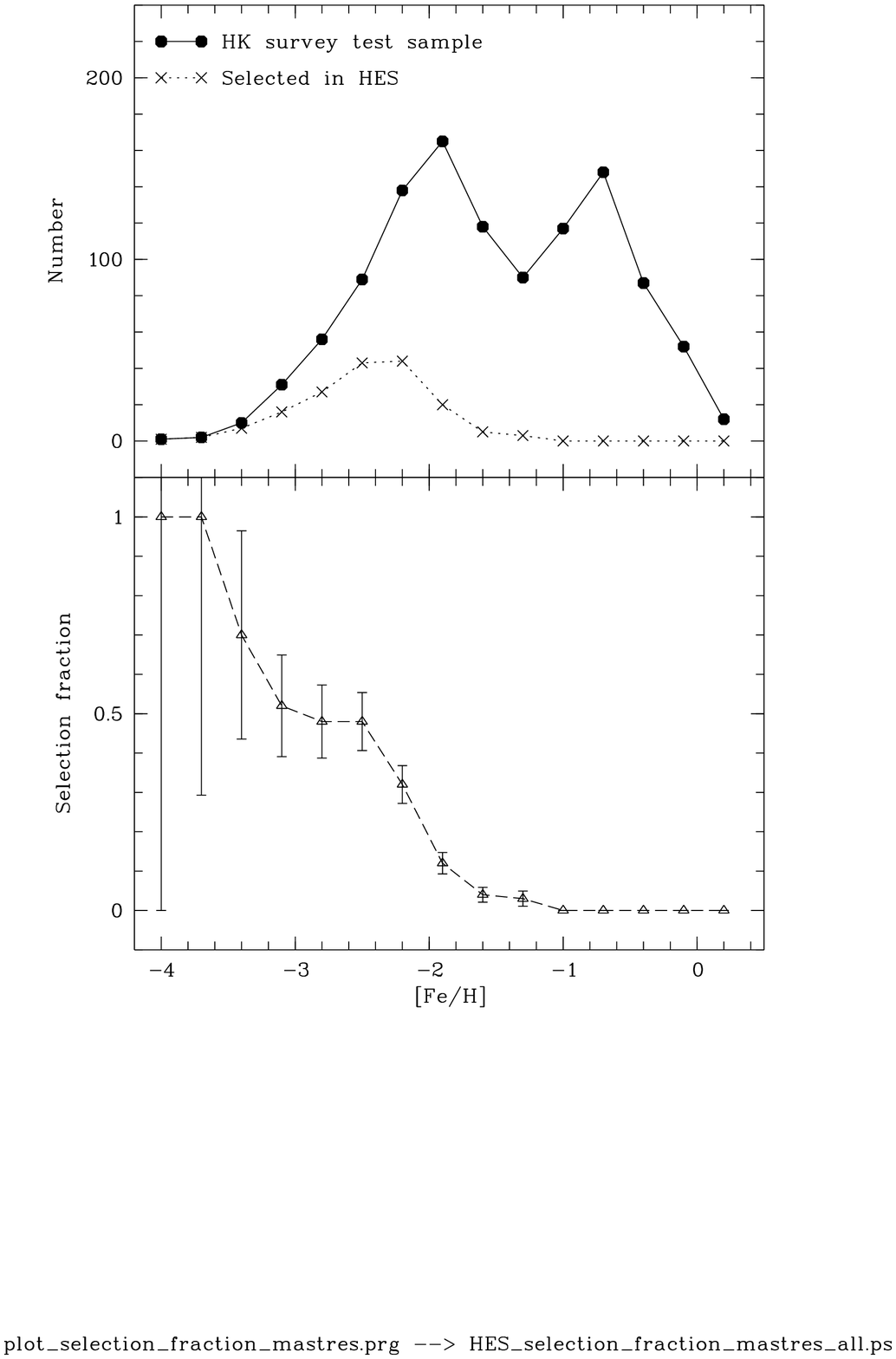}
  \caption{\label{Fig:HES_selection_fraction} HES selection fractions relative
    to a test sample of 1121 HK-survey stars present on HES plates.}
\end{figure}

We evaluated the selection of metal-poor candidates in the HES by means of a
test sample of 1121 HK survey stars present on HES plates for which [Fe/H] has
been determined based on moderate-resolution follow-up spectra, using an
updated version of the methods of \citet{Beersetal:1999}. The test sample has
been restricted to stars for which CCD $B$ and $V$ photometry exists, so that
the evaluation is based on the most reliable [Fe/H] estimates. Furthermore,
stars have been removed from the test sample if either the CCD $V$ magnitude
deviates by more than 1\,mag from the HES $V$ magnitude, or if the CCD $B-V$
color deviates from the HES $B-V$ color by more than 0.5\,mag. These stars are
very likely mis-identifications of HK survey stars, resulting from the fact
that a search box as large as $15\arcsec \times 15\arcsec$ was used during the
cross-identification, since the HK survey coordinates can have uncertainties
of more than 10\,arcsec. The result of the evaluation of the selection is
shown in Fig. \ref{Fig:HES_selection_fraction}.

In every candidate selection, there is a tradeoff between selection efficiency
and completeness. Relaxed selection criteria lead to high completeness but
lower efficiency (i.e., a higher number of false positives will contaminate
the sample), while strict criteria lead to low completeness but high
efficiency. Our evaluation of the selection shows that in the HES, a good
compromise has been reached. While 738 of the 764 test sample stars at
$\mathrm{[Fe/H]} > -2.0$ were rejected by the HES selection, resulting in a
low number of false positives, a satisfactory level of completeness at the
low-metallicity end is maintained: 20 of the 37 stars with $\mathrm{[Fe/H]} <
-3.0$ have been selected, and all four stars with $\mathrm{[Fe/H]} < -3.5$
have been recovered. Here, [Fe/H] refers to the value that was determined from
moderate-resolution follow-up spectra. These four stars are HE~0005$-$3547 $=$
CS~22876-032 \citep{Molaro/Castelli:1990}, with $\mathrm{[Fe/H]}=-3.7$
\citep{Norrisetal:2000}; the bright ($B_{\mbox{\scriptsize HES}}=12.9$) star
HE~0044$-$3755 $=$ {\cd} \citep{Bessell/Norris:1984} $=$ CS~22188-048, with
$\mathrm{[Fe/H]}=-4.0$ \citep{Norrisetal:2000,Francoisetal:2003}; HE~0305-5442
$=$ CS~22968-014, with $\mathrm{[Fe/H]}=-3.6$ \citep{Cayreletal:2004}; and
HE~2356-0410 $=$ CS~22957-027, with $\mathrm{[Fe/H]}=-3.4$
\citep{Norrisetal:1997b}. The quoted iron abundances were derived from
high-resolution spectra.

\begin{figure}[htbp]
  \centering
  \includegraphics[clip=true,bb=123 354 433 556,width=9cm]{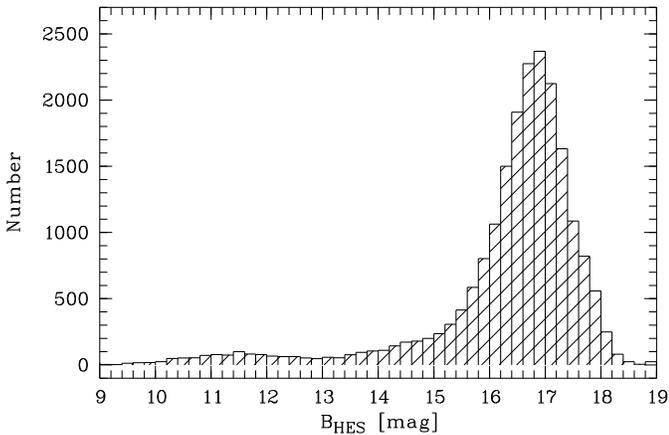}
  \caption{\label{Fig:mphscan_mag_distrib} $B$ magnitude distribution of the
  accepted HES candidates.}
\end{figure}

\section{Results}\label{Sect:Results}

Application of the selection criteria to all 379 HES fields yielded 26,928 raw
candidates, of which 5668 were rejected during the visual inspection. The
largest fraction of them are stars which are clearly too hot (i.e.,
hotter than the main-sequence turnoff of old, metal-poor halo stars; $\teffm
\sim 6800$\,K), as judged from the strength of the Balmer lines (1478 stars),
followed by sources having HES spectra corrupted by plate artifacts (693
objects) or overlapping spectra (574 objects). Another 947 raw candidates were
present on more than one plate quarter or on more than one field. The
corresponding 989 multiple entries in the candidate list were removed. In
total, we are left with 20,271 accepted candidates, which are listed in
Tab.~A.1. This corresponds to a surface density of 3 candidates per square
degree.

In Tab.~A.2 we provide the results of a cross-identification of the HES
candidates with the visually-selected candidates from the HK survey
(\citealt{BPSI,BPSII}; HK-I) as well as with the lists of high-proper-motion
stars of \citet{Ryan/Norris:1991b} and \citet{Carneyetal:1994}.

The magnitude distribution of the accepted HES candidates is shown in Fig.
\ref{Fig:mphscan_mag_distrib}. We define the magnitude limit of a survey as
the ``point of half height'' of the faint end of the magnitude distribution;
i.e., the magnitude at which the number of objects per magnitude bin has
dropped to half the maximum number of objects per magnitude bin of the
distribution. When adopting this definition, the magnitude limit of the HES
for metal-poor stars is $B = 17.2$. For comparison, the corresponding
magnitude limit of the HK survey, as judged from 2267 HK survey stars present
on HES plates, is $B = 15.2$.

Tab. \ref{Tab:candidate_statistics} summarizes the number of stars in each
candidate class, as well as the number of stars selected by each of the three
selection criteria. From these numbers we conclude that the KP/$(J-K)_0$
criterion is by far the most ``relaxed'' one, since in each of the four
candidate classes at least 70\,\% of the candidates have been selected by this
criterion.

\begin{table}[htbp]
 \centering
 \caption{Number of stars in each candidate class and selected by each of the
   three selection criteria. The sum of the numbers in rows 1--3 in each
   column disagree with the total listed in row 4, because stars can be
   selected by more than one criterion. }
 \label{Tab:candidate_statistics}
  \begin{tabular}{lrrrrr}\hline\hline
                   & \multicolumn{4}{c}{Candidate class} & \\
   \rb{Criterion}  & \texttt{mpca} & \texttt{unid} & \texttt{mpcb} & 
   \texttt{mpcc}   & \rb{Total} \\\hline
    KP/$(B-V)_0$       & $406$ & $1409$ & $4496$ & $2983$ &  $9294$\\
    KP/$(J-K)_0$       & $600$ & $1840$ & $7052$ & $6663$ & $16155$\\
    \ion{Ca}{ii}~K non-detection & $369$ & $1222$ & $2717$ & $1481$ &  $5789$\\\hline
    Total              & $772$ & $2533$ & $8872$ & $8094$ & $20271$\\\hline
  \end{tabular}
\end{table}

Preliminary results of the spectroscopic follow-up observations indicate that
the subset of candidates selected by the KP/$(J-K)_0$ criterion contains a
significantly higher number of false positives than the subsets selected by
the other criteria; in particular, a considerable number of stars that are
too blue. Tests suggest that this can largely be rectified by accepting only
candidates having $(J-K)_0 > 0.2$ \emph{and} $(B-V)_0 > 0.3$. It appears that
the use of the de-reddened $J-K$ color alone is not very efficient in
rejecting the bluer stars, probably due to the fact that the
1-$\sigma$ uncertainty of the 2MASS $K$ magnitudes typically exceeds
$0.1$\,mag for stars fainter than $V=16$.

\section{Conclusions}\label{Sect:Conclusions}

The high-quality digitization and accurate calibration of the photographic HES
plates made it possible to perform a quantitative selection of candidate
metal-poor stars, yielding 20,271 objects. The HK survey of Beers, Preston,
and Shectman paved the way for our efforts. In particular, we employed HK
survey stars present on HES plates for the calibration of the line indices and
the $B-V$ colors measured from HES spectra. For fixing the positions of the
selection cutoff lines in color versus KP index parameter space, we rely on
the techniques of \citet{Beersetal:1999} for determination of [Fe/H] from the
aforemention observables. The work of \citet{Beersetal:1999} in turn is based
on medium-resolution spectroscopy of thousands, and high-resolution
spectroscopy of hundreds of stars from the HK survey.

One of the advantages of a quantitative candidate selection is that the
selection function is well-defined. That is, for any point in the color versus
KP index parameter space, one can compute by means of simulations the
probability that an object would be selected as a candidate, given the
measurement uncertainties of the relevant observational quantities. A precise
knowledge of the selection function is important e.g. for statistical studies
like determining the halo MDF, because one needs to understand how the
candidate selection modifies the shape of the MDF. Such simulations, as well
as an attempt to determine the halo MDF from a sample of HES stars for which
follow-up observations already exists, are in progress (Sch\"orck et al., in
preparation).

Judging from a test sample of 1121 HK survey stars present on HES plates, the
candidate selection in the HES is very efficient in rejecting stars with
$\mathrm{[Fe/H]}>-2.0$, while at the same time the HES sample is highly complete
at the lowest metallicities. However, the result that 100\,\% of the test
sample stars with $\mathrm{[Fe/H]}<-3.5$ were recovered in the HES is based on
small number statistics (i.e., only 4 stars). Hence it is desirable to extend
the test sample in the future, e.g. with metal-poor stars to be identified in
the Southern Sky Survey \citep{Kelleretal:2007}.

Due to the fainter magnitude limit of the HES compared to the HK survey (i.e.,
$B=17.2$ compared to $B=15.2$), the survey volume for metal-poor stars was
extended by a factor of $\sim 10$ compared to the HK survey. Hence, it was
possible to identify in the HES objects which are too rare to be found with
the HK survey; e.g., stars at $\mathrm{[Fe/H]}<-5.0$.  The fainter HES limit
also allows one to explore outside the inner halo, which
\citet{Carolloetal:2007} have argued possesses a MDF that differs from that of
the outer halo. The inner halo, according to these authors, has a peak
metallicity ($\mathrm{[Fe/H]}=-1.6$) that is about a factor of three higher
than that of the outer halo ($\mathrm{[Fe/H]}=-2.2$).

In our spectroscopic follow-up observations we mainly focus on the best (i.e.,
candidates assigned to classes \texttt{mpca} and \texttt{unid}) and brightest
(i.e., $B<16.5$) candidates. We have also endeavored to complete the
observations in a few selected fields for statistical studies, such as
determining the shape of the low-metallicity tail of the halo MDF. The results
will be presented in future papers.

There are too many HES candidates to be processed in the course of our
collaborative efforts alone. We therefore encourage the community to obtain
follow-up spectroscopy of candidates that have not yet been observed by us.
A list of the candidates in need of observations is available on request from
the first author.

\begin{acknowledgements}
  We thank J.E. Norris for comments on an earlier version of this paper which
  resulted in considerable improvements. We are grateful to S.G. Ryan for
  providing us with an electronic version of Table~3 of
  \citet{Ryan/Norris:1991b}. This publication makes use of data products from
  the Two Micron All Sky Survey, which is a joint project of the University of
  Massachusetts and the Infrared Processing and Analysis Center/California
  Institute of Technology, funded by the National Aeronautics and Space
  Administration and the National Science Foundation. N.C. and D.R.
  acknowledge financial support from Deutsche Forschungsgemeinschaft through
  grants Ch~214/3 and Re~353/44. N.C. is a Research Fellow of the Royal
  Swedish Academy of Sciences supported by a grant from the Knut and Alice
  Wallenberg Foundation. A.F. is supported through the W.J. McDonald
  Fellowship of the McDonald Observatory. T.C.B. acknowledges partial funding
  for this work from grants AST~04-06784, AST~06-07154, AST~07-07776, and
  PHY~02-16873: Physics Frontier Center/Joint Institute for Nuclear
  Astrophysics (JINA), all awarded by the US National Science Foundation.
\end{acknowledgements}

\bibliographystyle{aa}
\bibliography{HES,mphs,ncastro,ncpublications,photometry,quasar,TimsRefs}
\newpage

\begin{appendix}

  \section{The catalog of candidate metal-poor stars}
  
  Tab. A.1 lists all 20,271 candidate metal-poor stars selected in the
  HES which were not rejected during the visual inspection. The table is made
  available only electronically. It contains the following columns:

  \begin{flushleft}
  \begin{tabular}{ll}
    hename        & HE designation\\
    HESid         & Unique HES identifier\\
    ra2000        & Right ascension at equinox 2000.0\\
    dec2000       & Declination at equinox 2000.0\\
    field         & ESO-SERC field number\\
    objtype       & Object type (\texttt{stars}/\texttt{bright})\\
    BHES          & Photographic $B$ magnitude\\ 
    BminV         & $B-V$ color, estimated from HES spectra\\
    UminB         & $U-B$ color, estimated from HES spectra\\
    EBminV        & $E(B-V)$\\
    BV0mphs       & $(B-V)_0$\\
    JminK0        & $(J-K)_0$\\
    VminK0        & $(V-K)_0$\\
    sn\_bj        & $S/N$ in $B_J$ band\\
    sn\_CaHK      & $S/N$ in Ca~H and K region\\
    KPHES         & KP index, measured in HES spectrum\\
    sigKP         & 1-$\sigma$ uncertainty of KP index\\
    GPHES         & GP index, measured in HES spectrum\\
    selCaKBminV0  & KP/$(B-V)_0$ selection flag\\
    selCaKJminK0  & KP/$(J-K)_0$ selection flag\\
    selnoCaK      & Ca~K non-detection selection flag\\
    canclass      & Candidate class (\texttt{mpca}/\texttt{unid}/\texttt{mpcb}/\texttt{mpcc})\\
    FEHK          & [Fe/H] estimate based on $(B-V)_0$ and KPHES\\
    FEHR          & [Fe/H] estimate based on $(J-K)_0$ and KPHES\\
    CFER          & [C/Fe], based on GPHES and KPHES\\
  \end{tabular}
  \end{flushleft}
  
  Comments on individual columns:
  \begin{description}
  \item[hename:] The Hamburg/ESO survey name consists of the letters ``HE'',
    the first four digits of the right ascension at equinox 1950.0, and the
    first four digits of the declination at equinox 1950.0.
  \item[HESid:] The average distance between two objects detected in the HES
    is 1.7\,arcmin, hence the HE name described above is often not capable of
    uniquely identifying HES sources; i.e., two or more HES sources have the
    same HE name. Therefore, an additional identifier was introduced, which
    consists of the first seven digits of the right ascension at equinox
    2000.0, the first six digits of the declination at equinox 2000.0, the
    plate quarter (a, b, c, or d), and the HES plate number. Example: For
    HE~0107$-$5240, which is located on quarter c of plate 2052 and at
    coordinates $\alpha=01^{\mathrm{h}}09^{\mathrm{m}}29\fs1$, $\delta=-52\degr
    24\arcmin 34\arcsec$, the HESid is HE0109291m522434c2052.
  \item[ra2000, dec2000:] The coordinates were derived from the DSS-I, and
    they are typically accurate to within 1\,arcsec.
  \item[BHES:] Photographic magnitude derived from DSS-I $B_J$ plates. For
    conversion between $B_J$ and Johnson $B$, we use the transformation $B =
    B_J + 0.28 \cdot (B-V)$, which is valid for main-sequence stars in the
    color range $-0.1<(B-V)<1.6$ \citep{Hewettetal:1995}. The $B-V$ color is
    the one derived from the HES spectra. A comparison of
    $B_{\mbox{\scriptsize HES}}$ with CCD $B$ magnitudes, using 963 HK survey
    stars present on HES plates and of object type \texttt{stars} or
    \texttt{bright}, yields that the 1-$\sigma$ uncertainty of
    $B_{\mbox{\scriptsize HES}}$ is $0.2$\,mag.
  \item[EBminV:] $E(B-V)$ color excess from the maps of
    \citet{Schlegeletal:1998}. Any reddening in excess of $E(B-V) = 0.10$ was
    reduced by 35\,\%, following \citep{Beersetal:2002}.
  \item[BV0mphs:] De-reddened $B-V$ color, where $B-V$ has been estimated from
    HES spectra using the calibrations described in Sect. \ref{Sect:BminVHES}.
  \item[JminK0:] The $J$ and $K$ magnitudes are from 2MASS.
  \item[VminK0:] $V$ was computed from the DSS-I $B$ magnitudes and the $B-V$
    color derived from HES spectra; $K$ is from 2MASS.
  \item[sn\_bj:] Average $S/N$ per pixel of the HES spectra in the $B_J$ band.
  \item[sn\_CaHK:] Average $S/N$ per pixel of the HES spectra in the Ca~H and K
    wavelength region.
  \item[FEHK, FEHR:] Estimates are given only if the \ion{Ca}{ii}~K line is
    detected at $> 1\,\sigma$ level in the HES spectrum.
  \end{description}
  
  Tab. A.2 (available electronically only) contains the names of 276 stars
  which were originally discovered in previous surveys for metal-poor stars
  and which were re-discovered in the HES.

  \begin{flushleft}
  \begin{tabular}{ll}
    HESid         & Unique HES identifier\\
    hename        & HE designation\\
    HKIname       & HK-I identifier\\
    othernames    & Other names\\
  \end{tabular}
  \end{flushleft}

\end{appendix}

\end{document}